\documentstyle[aps,epsf,twocolumn]{revtex}
\tighten
\begin{document}
\draft
\twocolumn[\hsize\textwidth\columnwidth\hsize\csname
@twocolumnfalse\endcsname


\title{Late time dynamics of scalar perturbations outside black holes.\\
                         II. Schwarzschild geometry}
\author{Leor Barack\cite{Email}}
\address {Department of Physics,
          Technion---Israel Institute of Technology, Haifa, 32000, Israel}
\date{\today}
\maketitle

\begin{abstract}
We apply a new analytic scheme, developed in a preceding paper,
in order to calculate the late time behavior of scalar test fields
evolving outside a Schwarzschild black hole.
The pattern of the late time decay at future null infinity
is found to be the same as in the shell toy-model studied in the
preceding paper.
A simple late time expansion of the scalar field is then used,
relying on the results at null infinity,
to construct a complete picture of the late time wave behavior
anywhere outside the black hole.
This reproduces the well known power-law tails at time-like
infinity and along the event horizon.
The main motivation for the introduction of the new approach
arises from its applicability to rotating black holes, as shall be
discussed in a forthcoming paper.
\end{abstract}


\vspace{6ex}
]

\section{introduction} \label{secI}

It is well established that the gravitational field of a
generically forming black holes relaxes at late time to a ``no-hair''
stationary Kerr-Newman geometry.
It was first demonstrated by Price \cite{Price72}, regarding
gravitational and electro-magnetic perturbations
of the Schwarzschild black hole (SBH) exterior, that the fields die
off at late time with an inverse power-law tail.
For a spherical-harmonic wave mode of multipole number $l$,
it was shown that a $t^{-(2l+2+p)}$ decay tail ($t$ being the Schwarzschild
time coordinate) will be detected by a static observer outside the
black hole, with $p=1$
if an initially compact perturbation is considered, or $p=0$
in case that a static field existed outside the central object before
the onset of collapse.

These results were later confirmed using several different techniques,
both analytic and numerical
\cite{Gundlach94I,Winicour94,Leaver86,Andersson97},
and were generalized to other spherically-symmetric spacetimes
\cite{Gundlach94I,Ching95,Burko97}.
The application of perturbative (linear) approaches is encouraged
by numerical analysis of the fully non-linear dynamics of the fields
\cite{Burko97,Gundlach94II}, which indicates virtually the same late
time pattern of decay, as for the minimally-coupled (linear)
fields.

Power-law decay tails are exhibited by fields at late time, because
in a curved spacetime waves do not propagate merely along light
cones, even when the fields are massless. Rather, the waves spread
inside the light cones due to scattering off spacetime curvature.
As suggested by previous studies, the late time
behavior of these waves is characteristic of merely the
large distance structure of spacetime.
This implies that the phenomenon of late time tails may not necessarily
be restricted to the exteriors of black holes.
For example, late time tails are found to form during the purely
spherical collapse of a self-gravitating minimally coupled scalar field,
even when the collapse fails to create a black hole
\cite{Gundlach94II}.
Conversely, no power law tails are detected in the
non-asymptotically-flat geometries of Schwarzschild--de Sitter and
Reissner--Nordsr\"{o}m--de Sitter black holes \cite{Brady97}
(instead, the field is found to die off exponentially at late time
in theses cases).

All previously mentioned studies were benefited from the
simplicity of spherical symmetry.
Yet, an astrophysically realistic model should clearly employ
a rotating central object.
Thus, apparently the most tempting generalization
of the analysis involves the inclusion of angular momentum in the
background geometry.
A first progress in this direction has been achieved recently with the
introduction of a full (1+2 dimensions) numerical analysis of wave
dynamics in Kerr spacetime, by Krivan {\em et al.}
\cite{Krivan96,Krivan97}.
So far, however, no {\em analytic} scheme has been proposed for the
investigation of wave dynamics in Kerr.

In a preceding paper (to be referred to as {\bf paper I}) we introduced
an analytic technique for the study of late time behavior of
fields in asymptotically-flat spacetimes. The prime motivation for the
introduction of the new scheme was its applicability to rotating
black holes.
To examine the essential features of the proposed calculation
scheme, we applied it in paper I to study the simple toy-model of
a scalar field evolving outside a spherically-symmetric thin shell of
matter.
In that case, the new technique, based on what we called ``{\em the
iterative expansion}'', allowed a simple and rigorous
derivation of the late time waves-form at null infinity.
In the present paper we apply a variant of the iterative scheme
in order to analyze the evolution of scalar waves on the background
of the {\em complete Schwarzschild geometry}. Again, this method
will enable the analytic calculation of the late time behavior at
null infinity.
We shall than show how, relying on the results at null infinity,
it becomes rather simple to construct a complete picture of the
late time decay anywhere outside the black hole, in particular
along the event horizon.

There are several reasons why we think it is worthwhile to first
analyze the already
well-studied case of a SBH, rather than directly focus on the
more interesting case of the Kerr black hole.
First, this will enable us to test our scheme against the
well-established results available in the Schwarzschild case.
Secondly, many parts of the analysis in Schwarzschild shall
later be directly employed when analyzing scalar waves in Kerr
\cite {Ours,Barack97}.
Finally, the analysis in Schwarzschild will appear to be valuable
on its own right, providing, in some respects, a more complete picture
of the late-time wave behavior than already available.

In the shell model, spacetime is flat at small distances (inside the
shell).
For that reason, the complete internal geometry could be exactly
accounted for by merely the ``Minkowski-like'' first component of
the iterative expansion (denoted in paper I by $\Psi_{0}$).
(we remind that the terms $\Psi_{N\geq 1}$ of the iterative
expansion were describing deviations from flat geometry,
namely curvature effects outside the shell.)
The complete Schwarzschild manifold, however, does not share this convenient
property, as in this case spacetime is highly curved at small distances.
This will enforce us to choose for another ``basis'' potential for the
iterative scheme at small $r$ (other than the purely-centrifugal potential
$V_{0}$ chosen in the framework of the shell model), and will thus somewhat
complicate the
technical details of the analysis. Nevertheless, the basic calculation
scheme, as well as the results at null infinity, shall remain
essentially the same as in the shell model.

\subsection*{arrangement of this paper}

This paper is arranged as follows. In sec.\ \ref{secII} we give a
mathematical formulation of the wave evolution problem in Schwarzschild
as a characteristic two-dimensional initial-value problem.
In sec.\ \ref{secIII} we introduce the iterative scheme to be used
to allow an analytic treatment of the mathematical problem.
We apply the iterative calculation scheme is sections \ref{secIV}
through \ref{secVII}, obtaining an expression for the waves-form
at late time at null infinity.
Then, in sec.\ \ref{secVIII}, a simple technique is applied to
obtain the late time behavior of the scalar field at any constant $r$
(including along the event horizon).
Sec.\ \ref{secIX} summarizes the results and discusses possible
extensions of the analysis.

\section{The initial value problem} \label{secII}
We consider the evolution of initial data, representing a generic pulse of
massless scalar radiation, on a fixed SBH background. The
scalar field is assumed to satisfy the (minimally-coupled) Klein--Gordon
equation
\begin{equation} \label{eq1}
\Phi_{;\mu}^{\ ;\mu}=0,
\end{equation}
where $\Phi$ represents the scalar wave.
The structure of spacetime affects the evolution of the scalar field through
the covariant derivatives, denoted in Eq.\ (\ref{eq1}) by semicolons.

Decomposing the field into spherical harmonics,
\begin{equation} \label{eq2}
\Phi(t,r,\theta,\varphi)=\sum_{l=0}^{\infty}\sum_{m=-l}^{l}\phi^{l}(t,r)
Y_{lm}(\theta,\varphi),
\end{equation}
we obtain an independent equation for each of the components
$\phi^{l}(t,r)$,
\begin{equation} \label{eq3}
f^{-1}(r) \phi_{,tt}^{l}-f(r)\phi_{,rr}^{l} -
\frac{2(r-M)}{r^{2}}\phi_{,r}^{l} + \frac{l(l+1)}{r^{2}}\phi^{l}=0.
\end{equation}
Here $t$, $r$, $\theta$ and $\varphi$ are the standard Schwarzschild
coordinates, $M$ is the mass of the black hole, $f(r)\equiv (1-2M/r)$,
and $l$ is the multipole number of the mode under consideration.

A more convenient form for the wave equation may be obtained in terms of a
new wave function
\mbox{$\Psi^{l}(t,r)\equiv r\phi^{l}(t,r)$}. To that end we
introduce the double-null (Eddington--Finkelstein) coordinates
$v\equiv t+r_{*}$ and $u\equiv t-r_{*}$,
where
\begin{equation} \label{eq3a}
r_{*}=r+2M\ln\left(\frac{r-2M}{2M}\right).
\end{equation}
The "tortoise" coordinate $r_{*}$ varies monotonically from $-\infty$
(the event horizon) to $+\infty$ (space-like infinity).

\begin{figure}[htb]
\input{epsf}
\centerline{\epsfysize 5cm \epsfbox{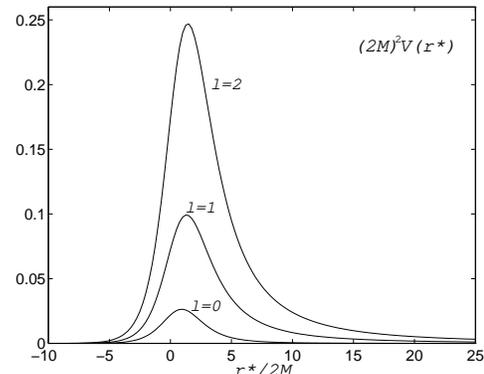}}
\caption{\protect\footnotesize Effective potential for scalar waves in
          Schwarzschild spacetime. With $r_{*}$ defined as in Eq.\
          (\ref{eq3a}), we have $r_{*}=0$ corresponding to $r\simeq 2.56M$.}
\label{fig1}
\end{figure}

The wave equation now reads
\begin{equation} \label{eq4}
\Psi_{,uv}^{l}+V^{l}(r)\Psi^{l}=0,
\end{equation}
in which
\begin{equation} \label{eq5}
V^{l}(r)=\frac{1}{4}\left(1-\frac{2M}{r}\right)
\left[\frac{l(l+1)}{r^{2}}+\frac{2M}{r^3}\right]
\end{equation}
is an effective potential, accounting for both
centrifugal and curvature effects.
This effective potential is sketched in figure
\ref{fig1} as a function of $r_{*}$ for the sample values $l=0,1,2$.

Note the following features of the effective potential (valid for all values
of $l$), which play an important rule in our analysis.
$V(r)$ is {\em localized}
(in a sense apparent in figure \ref{fig1}), forming an effective
potential barrier for the waves. At large distance, $V(r)$ is
dominated by the centrifugal potential (reflecting
asymptotic flatness), with curvature-induced deviations which die
off as $\sim r^{-3}$.
At small $r_{*}$ values, $V(r)$ dies off exponentially in $r_{*}/M$
towards the event horizon, making the potential effectively zero
inside the potential barrier.
Evidently, the late time behavior at null infinity is affected
mostly by the shape of the potential at large distance.
Conversely, the early evolution (e.\ g.\ the quasinormal ringing
stage), is strongly related to the fine details of the potential shape at
small $r$ values.

Since each of the spherical harmonics modes evolves separately,
we henceforth discuss the evolution of a single mode of arbitrary multipole
number $l$. The superscript `$l$' (denoting $l$-dependence)
will usually be suppressed for brevity.

\begin{figure}[htb]
\input{epsf}
\centerline{\epsfysize 6cm \epsfbox{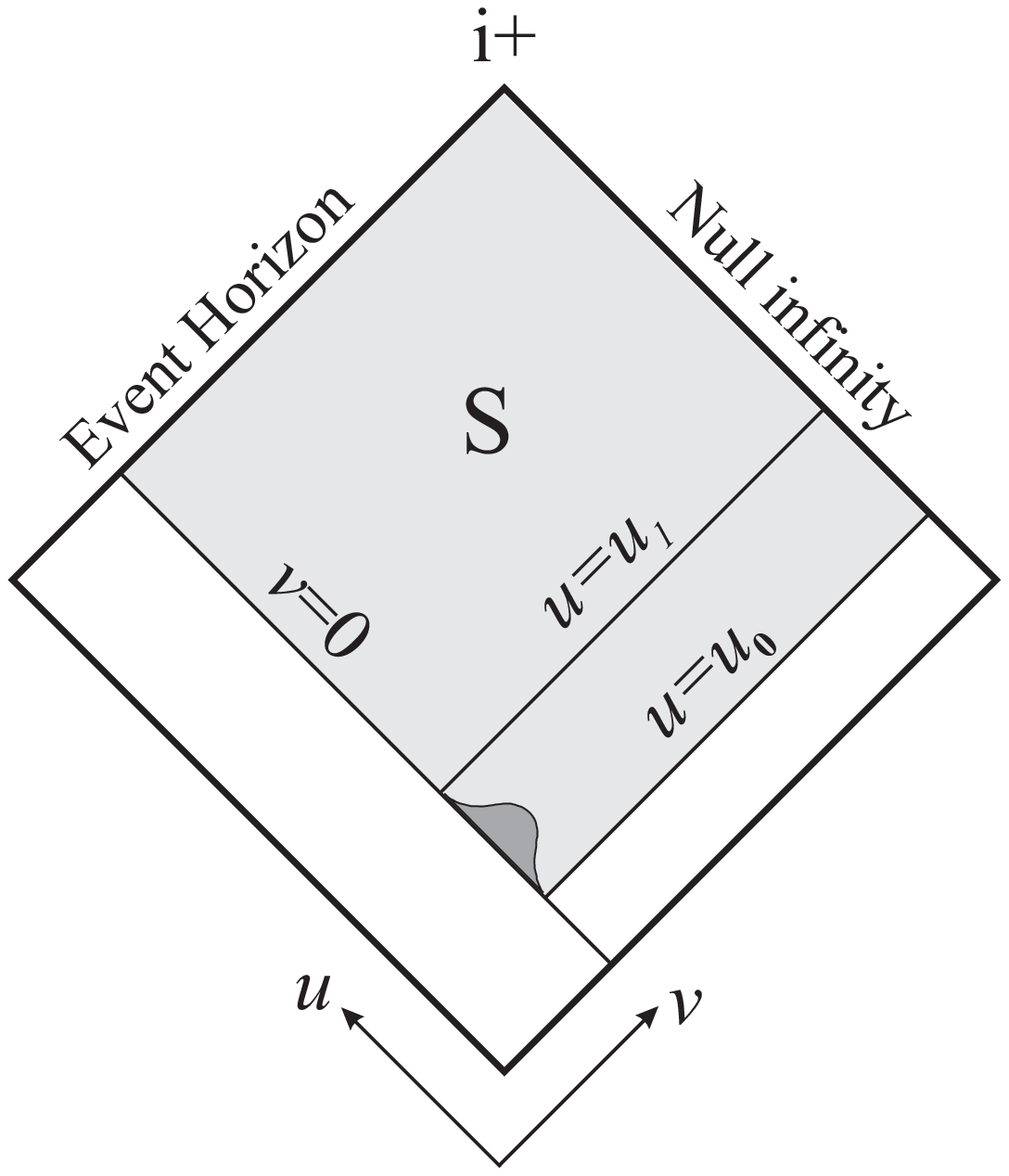}}
\caption{\protect\footnotesize The set-up of initial data. Shown is the
 Penrose diagram representing the external Schwarzschild geometry.
 The dark feature (artificially shown as if extended to $v>0$ values)
 represents the amplitude of some compact support initial function
 $\Gamma (u)$ on the ray $v=0$.
 The initial problem for the scalar field is well posed in region $S$
 (the shadowed area).}
\label{fig2}
\end{figure}

The initial data for the evolution problem shall be specified on two
characteristic (null) surfaces outside the event horizon,
as sketched in the Penrose diagram of figure \ref{fig2}.
We will first consider initial data in the form of some compact outgoing
pulse, specified on the ingoing null surface $v=0$\footnote{As long
as $u_{0}$ remains a free parameter in the analysis,
the specific choice $v=0$ causes no loss of generality, since spacetime is
time-translation invariant.},
\begin{equation} \label{eq6}
\left\{\begin{array}{l}
        \Psi(u=u_{0})=0 \\
        \Psi(v=0)=\Gamma(u)
        \end{array} ,
\right.
\end{equation}
where $\Gamma(u)$ is some function of a compact support (``the pulse'')
between retarded times $u=u_{0}$ and $u=u_{1}$.
As demonstrated in paper I for the shell model, the case of static
initial field can be later inferred in a simple way from the result
regarding a compact pulse.

The evolution equation (\ref{eq4}), supplemented by the initial conditions
(\ref{eq6}), establishes a well posed characteristic initial value problem
for the scalar field anywhere in the domain $S$ outside the event horizon
(see figure 2).
Since, manifestly, this problem poses no mathematical
irregularities, existence and uniqueness of a solution are
guaranteed by fundamental mathematical theory (see, for example,
\cite{Friedlander75}).

\section{The iterative expansion} \label{secIII}
To define the iterative expansion to be applied in the complete
SBH model, we first introduce a new parameter,
$r_{0}>0$, its value chosen so that $r(r_{*}=r_{0})$ is of order
$\gtrsim 2M$ [say, $r(r_{*}=r_{0})=3M$]. We then define
\begin{equation} \label{eq76}
V_{0}(r_{*})\equiv\left\{\begin{array}{ll}
                       0                        & ,r_{*}<r_{0} \\
                       \frac{l(l+1)}{4r_{*}^{2}} & ,r_{*}\geq r_{0}
                       \end{array}
                  \right.
\end{equation}
and
\begin{equation} \label{eq77}
\delta V(r)\equiv V-V_{0},
\end{equation}
in which $V(r)$ is the Schwarzschild effective potential given by
Eq.\ (\ref{eq5}).
The potential $V_{0}(r_{*})$ is so defined to account for the fact
that the actual effective potential is exponentially small at small
$r_{*}$ values. With this definition, the function $V_{0}(r_{*})$
approximates the form of the actual effective potential $V(r_{*})$
at both the very large and the very small values of $r_{*}$ (The
deviations, described by $\delta V$, become significant only at
intermediate distances, see figure \ref{fig1}).

Following the same procedure as in studying the shell model, we
define the {\em iterative expansion} by decomposing the scalar wave
$\Psi$ into an infinite sum,
\begin{equation} \label{eq77a}
\Psi=\sum_{N=0}^{\infty}\Psi_{N},
\end{equation}
in which the components $\Psi_{N}$ are
defined in a recursive way by the hierarchy of equations
\begin{equation} \label{eq78}
\Psi_{N,uv}+V_{0}\Psi_{N}=\left\{\begin{array}{ll}
                       0                     & ,N=0 \\
                       -(\delta V)\Psi_{N-1} & ,N>0
                       \end{array},
             \right.
\end{equation}
supplemented by the initial data
\begin{mathletters} \label{eq78a}
  \begin{equation} \label{eq78aa}
  \Psi_{N}(u=u_{0})= 0 \ \ \ \ (\forall N\geq 0),
  \end{equation}
     \begin{equation} \label{eq78ab}
     \Psi_{N}(v=0)=\left\{\begin{array}{ll}
                          \Gamma (u) & ,N=0 \\
                              0      & ,N>0
                          \end{array}.
                   \right.
     \end{equation}
\end{mathletters}
Formal summation over $N$ recovers the ``complete'' initial value
problem for the scalar wave $\Psi$.

It was indicated in paper I that in the analogous shell
model the iterative sum seems to converge rather efficiently
at late time to the actual field at null infinity, provided that the
initial pulse is specified at large distance.
In that case, it was suggested both numerically and analytically
that the ``complete'' wave is well approximated by merely the
function $\Psi_{1}$.
With this result in mind, we are going, in the following, to derive
exact analytic expressions for $\Psi_{0}$ and for the (time domain)
Green's function in the complete SBH model.
We shall then use these results to calculate the late
time form of $\Psi_{1}$ at null infinity in this model.

The above iterative expansion appears to be an effective
calculation scheme for all modes $l$ of the scalar radiation,
except for the monopole mode $l=0$. This is unlike the scheme used
for the shell model in paper I, which held equally-well for all
modes $l$ with no exception. The reason for this difference
between the two models in the monopole case will be discussed
later. The calculation to follow shall regard only the modes with $l>0$.

\section{Derivation of $\Psi_{0}$} \label{secIV}
We first obtain an explicit expression for $\Psi_{0}$, the first element of
the iterative expansion. Since the only discontinuity in the potential
function $V_{0}$ (at $r_{*}=r_{0}$) is bounded in magnitude, we
learn by
the wave equation (\ref{eq78}) that $\Psi_{0}$ and its first order
derivatives should be continuous anywhere. In the sequel we explicitly use
this fact in constructing an expression for $\Psi_{0}$.

\begin{figure}[htb]
\input{epsf}
\centerline{\epsfysize 6cm \epsfbox{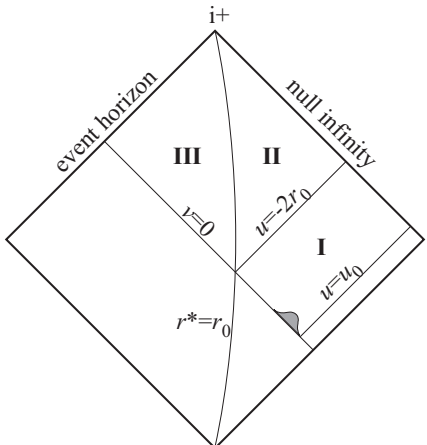}}
\caption{\protect\footnotesize Calculation of $\Psi_{0}$ in the complete
Schwarzschild geometry. The solution in each of the regions
labeled I, II, and III is discusses separately in the text.}
\label{fig3}
\end{figure}

We shall consider separately three distinct regions of the domain $S$,
as indicated in figure \ref{fig3}.
Regions I ($u_{0}<u<-2r_{0}$) and II ($u>-2r_{0}$,$r_{*}>r_{0}$) cover
the part of $S$ outside the surface $r_{*}=r_{0}$, while region III
$(r_{*}<r_{0})$ is the portion inside this surface.

A solution to Eq.\ (\ref{eq78}) for $N=0$ has the general form\footnote{
The most general solution at $r>r_{0}$
involves also an arbitrary function $h(v)$. However, for our
choice of initial setup (an outgoing initial pulse), the solution
$\Psi_{0}$ can be expressed in terms of a function
$g(u)$ solely. This issue is discusses in appendix A of paper I.}
\begin{mathletters} \label{eq79}
  \begin{equation} \label{eq79a}
  \Psi_{0I}=\sum_{n=0}^{l} A_{n}^{l}
            \frac{g_{0I}^{(n)}(u)}{(v-u)^{l-n}}
  \end{equation}
       \begin{equation} \label{eq79b}
       \Psi_{0II}=\sum_{n=0}^{l} A_{n}^{l}
                  \frac{g_{0II}^{(n)}(u)}{(v-u)^{l-n}}
       \end{equation}
            \begin{equation}\label{eq79c}
            \Psi_{0III}= F(u)+H(v) ,
            \end{equation}
\end{mathletters}
in which the labels I,II,III denote the region to which each specific
solution corresponds,
and where $g_{0I}(u)$, $g_{0II}(u)$, $F(u)$ and $H(v)$ are (yet)
arbitrary functions.
In the above equations the coefficients $A_{n}^{l}$ are given by
\begin{equation} \label{eq79d}
A_{n}^{l}=\frac{(2l-n)!}{n!(l-n)!},
\end{equation}
and the parenthetical indices indicate the number of times the functions
are differentiated.

Causality implies that in region I, the solution $\Psi_{0I}(u)$
cannot be sensitive to the form of $V_{0}$ at $r_{*}<r_{0}$. Thus it
must be identical to the solution derived in the shell model,
with the function $g_{0I}(u)$ explicitly related to the initial
data function by
\begin{equation} \label{eq80a}
g_{0I}(u)=\frac{1}{(l-1)!}\int_{u_{0}}^{u}\!
\left(\frac{u}{u'}\right)^{l+1}(u-u')^{l-1}\Gamma(u')du'
\end{equation}
(see paper I for details).

Now, with the initial condition $\Psi_{0III}(v=0)=0$\footnote{
Since we shall be interested mostly in the case where the initial pulse
is specified at large distance, we assume here that its support is confined
to the exterior of the sphere $r_{*}=r_{0}$.},
Eq.\ (\ref{eq79c}) implies that $\Psi_{0}^{III}$ is a function of $v$ only.
We can then use the continuity of $\Psi_{0,u}$ at $r_{*}=r_{0}$ to
derive a closed differential equation for the function
$g_{0II}(u)$:
\begin{equation} \label{eq81}
\frac{\partial}{\partial u}\left.\left\{
\sum_{n=0}^{l} A_{n}^{l} \frac{[g_{0II}(u)]^{(n)}}{(v-u)^{l-n}} \right\}
\right|_{r_{*}=r_{0}}=0 .
\end{equation}
This equation may be put into the form
\begin{equation} \label{eq82}
\sum_{n=0}^{l+1} B_{n}^{l} r_{0}^{n}[g_{0II}(u)]^{(n)}=0,
\end{equation}
in which the coefficient $B_{n}^{l}$ are given by
\begin{equation} \label{eq83}
B_{n}^{l}=\frac{2^{n}(2l-n)!}{n!(l-n+1)!}.
\end{equation}
Thus $g_{0II}(u)$ is a solution of a constant-coefficients
linear equation of order $l+1$. It therefore admits the form
\begin{equation} \label{eq84}
g_{0II}(u)=\sum_{i=1}^{l+1} C_{i} \exp(-\kappa_{i} u/r_{0}),
\end{equation}
where $C_{i}$ are constants, and the $l+1$ complex numbers $\kappa_{i}$
are the roots of the algebraic equation
\begin{equation} \label{eq85}
\sum_{n=0}^{l+1} B_{n}^{l} (-\kappa)^{n}=0 .
\end{equation}
The only properties of the numbers $\kappa_{i}$ important for our
discussion are that (i) these numbers are all distinct (for any given
value of $l$),
and that (ii) we have $Re(\kappa_{i})>0$ for all values of $l$ and $i$.
Hence $g_{0}$ (and also $\Psi_{0}$ itself) falls off exponentially
at late retarded time $u$.

The coefficients $C_{i}$ are determined by imposing continuity on
$\Psi_{0}$
at $u=-2r_{0}$, namely by requiring $(g_{0II})^{(j)}=(g_{0I})^{(j)}$
at $u=-2r_{0}$, for all $0\leq j\leq l$ .
This leads to a set of $l+1$ algebraic equations for the $l+1$
coefficients $C_{i}$, having the form
\begin{equation} \label{eq85a}
\sum_{i=1}^{l+1}M_{ji}C_{i}=\left. r_{0}^{j}g_{0I}^{(j)}\right|_{u=-2r_{0}}
\end{equation}
for $0\leq j\leq l$, where $M_{ji}\equiv (-\kappa_{i})^{j}\exp(2\kappa_{i})$.
In a matrix form, we have
\begin{equation} \label{eq85b}
\det \bbox{M}=\exp[2(\kappa_{1}+\cdots +\kappa_{l+1})]\det \bbox{K},
\end{equation}
where $\bbox{K}$ is the Vandermonde matrix
\begin{equation} \label{eq85c}
\bbox{K}=\left( \begin{array}{cccc}
    1             &        1          &  \cdots  &         1         \\
 -\kappa_{1}      &  -\kappa_{2}      &  \cdots  &  -\kappa_{l+1}     \\
  \vdots          &  \vdots           &          &  \vdots           \\
 -\kappa_{1}^{l}  &  -\kappa_{2}^{l}  &  \cdots  &  -\kappa_{l+1}^{l} \\

          \end{array}
  \right) ,
\end{equation}
which is always non-singular, provided only that the numbers $\kappa_{i}$
are all distinct (which is the case here).
Therefore the set of equations (\ref{eq85a}) has a unique
solution for the coefficients $C_{i}$.

We finally obtain $\Psi_{0II}$ by substituting for $g_{0II}$
in Eq.\ (\ref{eq79b}). This yields an expression of the form
\begin{equation} \label{eq86}
\Psi_{0II}=r_{*}^{-l} \sum_{n=0}^{l} \sum_{i=1}^{l+1} \alpha_{ni}
           \left(\frac{r_{*}}{r_{0}}\right)^{n} E_{i}(u),
\end{equation}
in which $\alpha_{ni}$ are constant coefficients (being certain
$l$-dependent functionals of the initial-data function $\Gamma(u)$),
and where the functions
\begin{equation} \label{eq86a}
E_{i}(x)\equiv \exp[-\kappa_{i} x/r_{0}]
\end{equation}
die off exponentially with respect to their argument for all
$1\leq i\leq l+1$.\footnote{
        In Eq.\ (\ref{eq86}), as well as in all other expressions
        for the various functions $\Psi$ to be appeared in this
        paper, it is to be understand that only the {\em real} (or
        alternatively---the {\em imaginary}) part is taken into
        account. The indication ``Re'' shall be omitted for
        brevity.}
We still have to derive an expression for $\Psi_{0}$ in region III,
that is at $r_{*}<r_{0}$.
By the continuity of $\Psi_{0}$ at $r_{*}=r_{0}$ we have
$\Psi_{0}^{III}(v)=\Psi_{0}^{II}(u\!=\!v\!-\!2r_{0})$.
It follows that
\begin{equation} \label{eq87}
\Psi_{0III}=r_{0}^{-l}\sum_{i=1}^{l+1}\alpha_{i}
                 E_{i}(v-2r_{0}),
\end{equation}
where $\alpha_{i}\equiv\sum_{n=0}^{l}\alpha_{ni}$, and where the
functions $E_{i}$ are those defined in Eq.\ (\ref{eq86a}).

\begin{figure}[htb]
\input{epsf}
\centerline{\epsfysize 6cm \epsfbox{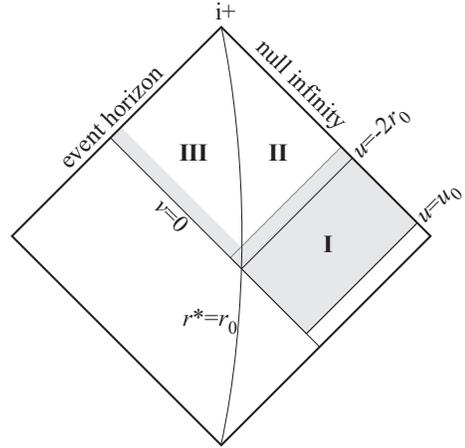}}
\caption{\protect\footnotesize Domain of the function $\Psi_{0}$
in Schwarzschild. Dark colored areas indicate regions where $\Psi_{0}$ is
not exponentially small. Note that $\Psi_{0}$ ``penetrates'' the potential
barrier only through a narrow ingoing ray (of typical width
$\sim 2r_{0}$).}
\label{fig4}
\end{figure}

We conclude that $\Psi_{0}$ ``penetrates'' the potential barrier only
through a narrow null ray of typical width $\sim 2r_{0}$ adjacent to the
initial ingoing ray $v=0$.
It has significant amplitude only in a
``main'' region $u_{0}<u\lesssim 0$ and along that penetrating null ray.
Elsewhere, $\Psi_{0}$ is found to be exponentially
small (in retarded time $u$ at $r_{*}>r_{0}$ or in advanced time $v$ at
$r_{*}<r_{0}$) . This result (valid for all $l\geq 1$) is illustrated
in figure \ref{fig4}.

\section{Construction of the Green's function} \label{secV}
In this section we derive an analytic expression for the (retarded) Green's
function corresponding to the operator
$\partial_{u}\partial_{v}+V_{0}(r_{*})$,
with $V_{0}(r)$ defined in Eq.\ (\ref{eq76}).
The (retarded) Green's function $G(u,v;u',v')$ is defined by the
equation
\begin{equation} \label{eq87a}
G,_{uv}+V_{0}G=\delta(u-u') \delta(v-v'),
\end{equation}
supplemented by the causality condition $G(v<v')=G(u<u')=0$, where $(u',v')$
are the null coordinates of a scalar ``point'' source (in the 1+1 dimensions
representation), and $(u,v)$ is where we evaluate the field this source
induces.
[It will become evident by construction that this condition
specifies a unique solution to Eq.\ (\ref{eq87a}).]
In view of the results obtained for
$\Psi_{0}$, we shall have to consider both `external' ($r_{*}'>r_{0}$) and
`internal' ($r_{*}'<r_{0}$) sources. In what follows we treat each of these
two cases separately.

\subsection {External sources} \label{subsecV1}
\begin{figure}[htb]
\input{epsf}
\centerline{\epsfysize 6cm \epsfbox{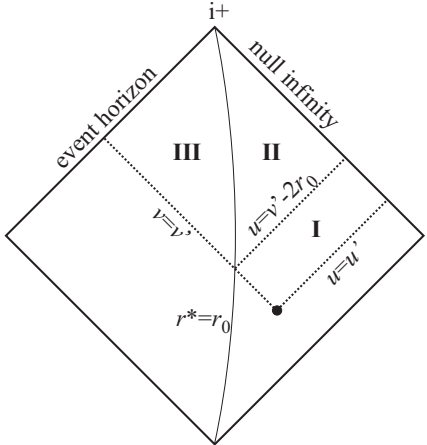}}
\caption{\protect\footnotesize Construction of the Green's function for
a scalar source sphere at $r_{*}'>r_{0}$. The three regions I, II,III,
defined with respect to that source, are
treated separately in the text.}
\label{fig5}
\end{figure}

We first consider a ``point'' source
located at null coordinates $(u',v')$ outside the surface
$r_{*}=r_{0}$ (thus $v'-u'>2r_{0}$). For this fixed source, we look for
the Green's function at any evaluation point $(u,v)$.
To that end we separate the future light cone of the point source into
three regions, as indicated in figure \ref{fig5}. Regions I and II
correspond to evaluation points outside the surface $r_{*}=r_{0}$, while
region III corresponds to internal evaluation points.

We first observe that in Region I (that is at $u<v'-2r_{0}$) the
Green's function cannot depend on the form of the potential at $r_{*}<r_{0}$
(as implied by causality), and thus in this region is must be the same as in
the shell model (outside the shell).
Therefore, by Eqs.\ (33) and (35) of paper I we find that the Green's
function in region I reads
\begin{equation} \label{eq87b}
G^{I}(u,v;u',v') = \sum_{n=0}^{l} A_{n}^{l}
                   \frac {[g_{G}^{I}(u;u',v')]^{(n)}}{(v-u)^{l-n}},
\end{equation}
in which the differentiation is with respect to $u$, $A_{n}^{l}$ are the
coefficients given in Eq.\ (\ref{eq79d}), and
\begin{equation} \label{eq33}
g_{G}^{I}(u;u',v')=\frac{1}{l!}
                   \left[\frac {(v'-u)(u-u')}{(v'-u')}\right]^{l}.
\end{equation}

Now, in regions II and III Eq.\ (\ref{eq87a}) is homogeneous, hence
the solutions for the Green's function in these two regions are of the form
\begin{mathletters} \label{eq88}
  \begin{equation}\label{eq88a}
  G^{II}=\sum_{n=0}^{l} A_{n}^{l}
         \frac {[g_{G}^{II}(u)]^{(n)}}{(v-u)^{l-n}}\\
  \end{equation}
      \begin{equation}\label{eq88b}
       G^{III}=G^{III}(v) ,
      \end{equation}
\end{mathletters}
where the functions $g^{II}(u)$ and $G^{III}(v)$ are yet to be
determined.

By analogy with Eq.\ (\ref{eq84}) we then have
\begin{equation} \label{eq89}
g_{G}^{II}(u)=\sum_{i=1}^{l+1} \bar{C}_{i}(u',v')
              \exp(-\kappa_{i} u/r_{0}),
\end{equation}
with $\kappa_{i}$ being the same numbers as in Eq.\ (\ref{eq84}),
and
where the $l+1$ coefficients $\bar{C}_{i}(u',v')$ are to be determined such
that the Green's function is continuous along the ray $u=v'-2r_{0}$.
This requirement leads to a set of $l+1$ equations for the coefficients
$\bar{C}_{i}(u',v')$, reading
\begin{equation} \label{eq89a}
\sum_{i=1}^{l+1}\bar{M}_{ji}\bar{C}_{i}=\left. r_{0}^{j}[g_{G}^{I}(u)]^{(j)}
\right|_{u=v'-2r_{0}}
\end{equation}
(for $0\leq j\leq l$),
where $\bar{M}_{ji}\equiv (-\kappa_{i})^{j}\exp[-\kappa_{i}(v'-2r_{0})
/r_{0}]$.
The solution (which always exists) is
\begin{equation} \label{eq89b}
\bar{C}_{i}=\exp[\kappa_{i}(v'-2r_{0})/r_{0}]
       \sum_{j=0}^{l}\left. K^{-1}_{ij}r_{0}^{j}[g_{G}^{I}(u)]^{(j)}
       \right|_{u=v'-2r_{0}} ,
\end{equation}
with $K^{-1}_{ji}$ being the elements of the matrix reciprocal to the
Vandermonde matrix (\ref{eq85c}).
Inserting the explicit expression for $g_{G}^{I}$ and using
Eq.\ (\ref{eq88a}), we can finally obtain for the Green's function in
region II,
\begin{equation} \label{eq94}
G^{II}\!=\!\!\!\sum_{n,j=0}^{l} \!\sum_{i=1}^{l+1}\beta_{nji}
       \frac{(r_{*}'-r_{0})^{l-j}(r_{0})^{l+j-n}}{(r_{*}')^{l}r_{*}^{l-n}}
       E_{i}(u-v'+2r_{0}),
\end{equation}
in which $r'_{*}\equiv (v'-u')/2$, and where $\beta_{nji}$ are
certain constant coefficients (depending on $l$ only).
(Recall that the functions $E_{i}$
die off exponentially with respect to their argument for all $i$.)

To obtain the Green's function in region III, we simply notice that
$G^{III}(v)=G^{II}(u=v-2r_{0})$, implied by the continuity of $G$ at
$r_{*}=r_{0}$.
It follows that
\begin{equation} \label{eq95}
G^{III}=\sum_{j=0}^{l} \sum_{i=1}^{l+1}\beta_{ji}
        \frac{(r_{*}'-r_{0})^{l-j}(r_{0})^{j}}{(r_{*}')^{l}}
        E_{i}(v-v'),
\end{equation}
where $\beta_{ji}\equiv\sum_{n=0}^{l}\beta_{nji}$.
(It is straightforward to verify that with this result, we have
$G^{III}(v=v')=1$ as necessary.)

\subsection {Internal sources} \label{subsecV2}
\begin{figure}[htb]
\input{epsf}
\centerline{\epsfysize 6cm \epsfbox{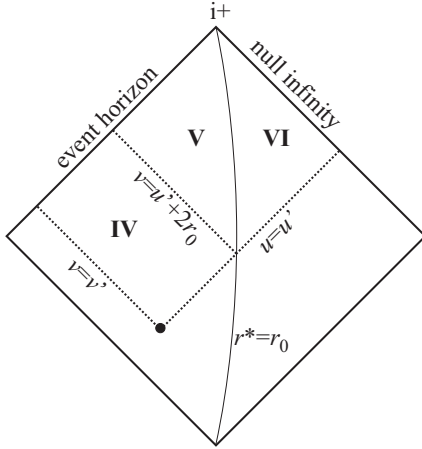}}
\caption{\protect\footnotesize Construction of the Green's function for
a scalar source sphere at $r_{*}'<r_{0}$. The three regions IV, V, VI are
treated separately in the text.}
\label{fig6}
\end{figure}

To obtain the Green's function for a source point located at $r_{*}'<r_{0}$,
we refer to figure \ref{fig6}, where again we indicate three regions,
defined with respect to a given source at $(u',v')$. Again we discuss
the construction of the Green's function in each of these regions in
separate.

In region IV we have $G^{IV}_{,uv}=\delta(v-v')\delta(u-u')$ by definition,
which (by causality) leads to
\begin{equation} \label{eq96}
G^{IV}=\theta(v-v')\theta(u-u'),
\end{equation}
with $\theta$ denoting the usual step function.

In region V the Green's function satisfies the homogeneous equation
$G^{V}_{,uv}=0$. With the continuity requirement $G^{V}(v=u'+2r_{0})=1$,
this means that $G^{V}=G^{V}(v)$.
Now, $G^{VI}$ is given in terms of a function $g_{G}^{VI}(u)$, in a way
analogous to $G^{II}$ in Eq.\ (\ref{eq88a}). By the continuity of $G_{,u}$
at $r_{*}'=r_{0}$ we must have $G^{VI}_{,u}(r_{*}=r_{0})=0$,
which is a linear
differential equation of order $l+1$ for the function $g_{G}^{VI}(u)$.
The solution is (in analogy to Eq.\ (\ref{eq89}),
\begin{equation} \label{eq97}
g_{G}^{VI}(u)=\sum_{i=1}^{l+1} \tilde{C}_{i}(u',v')
\exp[-\kappa_{i} u/r_{0}],
\end{equation}
with $\tilde{C}_{i}$ being certain coefficients.

To construct the coefficients $\tilde{C}_{i}$, we
match the function $G^{VI}$, as inferred by Eq.\ (\ref{eq97}), to
its value on the ray $u=u'$. This value may be deduced independently by
inserting the form $G^{IV}(u,v)=\bar{G}^{IV}(u,v)\theta(u-u')$ (implied
by causality) into equation (\ref{eq87a}), and observing
that a solution
must admit $\bar{G},_{v}=0$ along $u=u'$. This means that $G$ is
constant along this ray. By Eq.\ (\ref{eq96}) (and requiring continuity)
we then learn that this constant is unity.
Requiring $G^{VI}(u=u')=1$ for all $v$ then leads to
\begin{equation} \label{eq98}
\left\{\begin{array}{ll}
             [g_{G}^{IV}]^{(n)}(u')=0     &   , \ \ (0\leq n\leq l-1) \\
             {[}g_{G}^{IV}]^{(l)}(u')=1   &
      \end{array}.
\right.
\end{equation}
With Eq.\ (\ref{eq97}) , this constructs a set of $l+1$ linear algebraic
equations for the coefficients $\tilde{C}_{i}$. The solution reads
\begin{equation} \label{eq99}
\tilde{C}_{i}=r_{0}^{l} K^{-1}_{i,l+1} \exp[\kappa_{i}u'/r_{0}],
\end{equation}
where the numbers $\kappa_{i}$ are the same as for the external source.
(Recall that the matrix $\bbox{K}$ is always non-singular, hence
this solution exist and is unique.)

Using the results (\ref{eq97}) and (\ref{eq99}) we can finally obtain
\begin{equation} \label{eq100}
G^{VI}=\sum_{n=0}^{l} \sum _{i=1}^{l+1}\gamma_{ni}
       \left(\frac{r_{0}}{r_{*}}\right)^{l-n} E_{i}(u-u'),
\end{equation}
with the functions $E_{i}$ defined in Eq.\ (\ref{eq86a}), and where
$\gamma_{ni}$ are certain constant coefficients (depending only on $l$).

To obtain $G^{V}$, we simply notice that $G^{V}(v)=G^{VI}(u=v-2r_{0})$
(inferred by the continuity of the Green's function), hence
\begin{equation} \label{eq101}
G^{V}=\sum_{i=1}^{l+1} \gamma_{i} E_{i}(v-u'-2r_{0}),
\end{equation}
where $\gamma_{i}\equiv \sum_{n=0}^{l}\gamma_{ni}$.

\subsection {Fixed external evaluation point} \label{subsecV3}
\begin{figure}[htb]
\input{epsf}
\centerline{\epsfysize 6cm \epsfbox{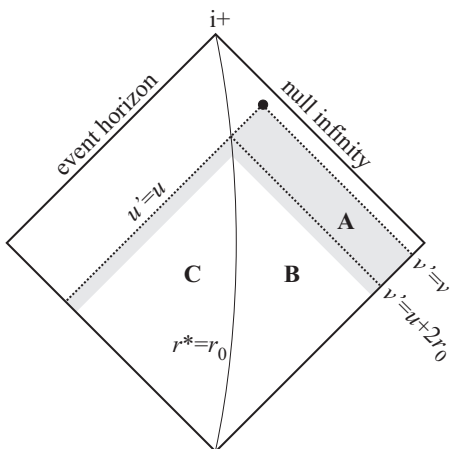}}
\caption{\protect\footnotesize Construction of the Green's function for
a given external evaluation point at $u,v$. Regions where the Green's
function is {\em not} exponentially small are indicated by dark color.}
\label{fig7}
\end{figure}

Thus far we considered the Green's function for a given sources at $(u',v')$,
as a function of the evaluation coordinates $(u,v)$. In practice, we shall
be interested in calculating the function $\Psi_{1}$ at a given location
(specifically---at
null infinity, for $u\gg M$), what will involve integration over all
possible sources. This requires knowledge of the form of the Green's
function at the evaluation location, as a function of the sources locations.
To that end we only need to re-interpret our previous
results: The expressions we have derived for the Green's function shall be
regarded as functions of the source coordinates $(u',v')$, with fixed
evaluation coordinates $(u,v)$. This reversed presentation of the results is
illustrated in figure \ref{fig7}. Indicated in this figure are the regions
of spacetime  in which scalar
sources influence the behavior of the scalar field at a fixed evaluation
point (with null coordinates $u,v$) outside the surface $r_{*}=r_{0}$.
Dark-colored areas in this figure indicate source regions where the
Green's function is {\em not}
exponentially-small, as inferred by Eqs.\ (\ref{eq87b}), (\ref{eq94}) and
(\ref{eq100}).

In region A ($u'\leq u$, $u+2r_{0}\leq v'\leq v$) $G$ is given by
Eqs.\ (\ref{eq87b}) and (\ref{eq33}), and is the same as in the shell model.
In region B ($v'\leq u+2r_{0}$, $r_{*}'\geq r_{0}$)
$G$ is given by Eq.\ (\ref{eq94}). It vanishes here exponentially towards
early advanced time $v'$, and it possesses significant amplitude only
within a narrow ingoing null ``band'' (of typical width $2r_{0}$),
adjacent to region A. Then, in region C ($u'\leq u$, $r_{*}'\leq r_{0}$)
the Green's
function [given by Eq.\ (\ref{eq100})] vanishes exponentially towards early
retarded time $u'$, and is of significant amplitude only within a narrow
outgoing null ``band'' (of typical width $2r_{0}$) adjacent to the ray
$u'=u$.

Like in the shell model, we find that the main region of effective sources
covers only the range $u\leq v'\leq v$ of advanced times.
In the shell model, however, the Green's function vanishes
identically outside this range (due to the divergent potential at the center
of symmetry), whereas in the complete SBH model it dies off exponentially
(and also ``penetrates'' trough the finite potential barrier at
$u'=u$).

\section{Calculation of $\Psi_{1}$ at null infinity} \label{secVI}
We have shown that in the complete SBH model, $\Psi_{0}$ gives only
an exponentially decaying contribution to the late time radiation. In this
section we calculate the contribution of $\Psi_{1}$ to this radiation
at null infinity, and show that it is characterized by the same power-law
tail of decay that was indicated in the shell model. Moreover, we show that
even the amplitudes of the waves are the same on both models, provided that
we choose $|u_{0}|\gg 2M\simeq r_{0}$ (the difference is of order
$r_{0}/u_{0}$).

\begin{figure}[htb]
\input{epsf}
\centerline{\epsfysize 6cm \epsfbox{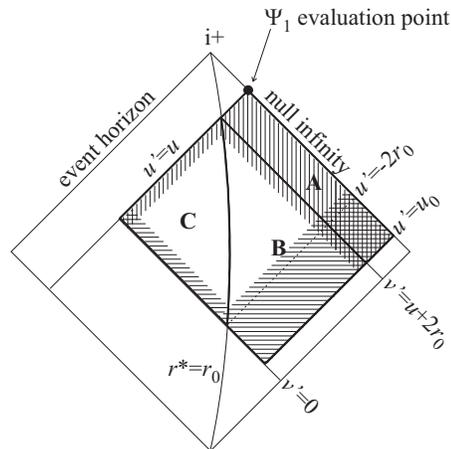}}
\caption{\protect\footnotesize calculation of $\Psi_{1}$ in the complete
Schwarzschild model. The region filled with horizontal lines is where the
amplitude of $\Psi_{0}$ is significant, while the region marked with vertical
lines is where the Green's function is significant, in accordance with the
discussion given in the text. Potentially, significant contribution to
$\Psi_{1}$ is expected to arise only from areas where the two indicated
regions overlap.}
\label{fig8}
\end{figure}

Before we present the detailed calculation of $\Psi_{1}$, we first give
some heuristic arguments concerning the expected results.
Figure \ref{fig8} shows the region
of spacetime in which scalar sources affect the behavior of the wave at null
infinity, at a given retarded time $u\gg M$. Also shown, superposed, is the
region where sources due to $\Psi_{0}$ exist.
Outside the overlapping of these two areas, the Green's function, or
$\Psi_{0}$, or both, are exponentially small. We expect
(and later show analytically) that sources outside the overlapping area
shall give only an exponentially decaying contribution to $\Psi_{1}$
at null infinity as $u\rightarrow \infty$.
We may thus focus only on the two overlapping regions shown in the
figure.
One of these regions lies inside the surface $r_{*}=r_{0}$ (see the figure).
It is
of ``dimensions'' $r_{0}\times r_{0}$, and is located near
$r_{*}=(-u/2)\ll M$.
In this location the potential function $V(r_{*})$ is exponentially small
(see figure \ref{fig1}), and thus the contribution from this area should be
exponentially small as well. We are left with the contribution of
sources at the ``main'' region [namely $u_{0}\leq u'\leq 0$, $u\leq
v'\leq v$], in which both the Green's function and $\Psi_{0}$ are having the
same form as in the shell model, except for in a narrow band (of width
$\sim 2r_{0}$) at the edge of this region. This suggests that for $r_{0}$
small enough, the calculation of $\Psi_{1}$ should yield a result very
close to that obtained in the shell model.

To confirm the above heuristic indications, we shall now calculate
$\Psi_{1}$ at null infinity.
In terms of the Green's function derived above, we formally have
\begin{equation} \label{eq102}
\Psi_{1}^{\infty}(u)=
\!-\!\int_{u_{0}}^{u}\!\!\!\!du'\!\!\int_{0}^{v}\!\!\!\!dv'
\,G^{\infty}(u;u',v')\delta V (u',v') \Psi_{0}(u',v'),
\end{equation}
where $\Psi_{1}^{\infty}$ and $G^{\infty}$ stand for the value of $\Psi_{1}$
and $G$ at null infinity (that is, for $v\rightarrow \infty$).
For the purpose of calculation, we separate the domain of integration
into three regions,
\begin{eqnarray} \label{eq103}
\lefteqn{\int_{u_{0}}^{u}\!\!\!du'\int_{0}^{v}\!\!\!dv' = }    \nonumber\\
 & &\int_{u+2r_{0}}^{\infty}\!\!\!\!\!\!\!\!\!dv'
                                                  \int_{u_{0}}^{u}\!\!\!du'+
    \int_{0}^{u+2r_{0}}\!\!\!\!\!\!\!\!\!\!\!dv'
                          \int_{u_{0}}^{v'-2r_{0}}\!\!\!\!\!\!\!\!\!\!\!du'+
    \int_{0}^{u+2r_{0}}\!\!\!\!\!\!\!\!\!\!\!dv'
                              \int_{v'-2r_{0}}^{u}\!\!\!\!\!\!\!\!\!du'
\end{eqnarray}
to be labeled {\bf A}, {\bf B}, and {\bf C}, respectively, as indicated
in figure \ref{fig8}.
In what follows we consider separately
the contribution from each of these three regions to $\Psi_{1}^{\infty}$.
We show that the contribution of region {\bf A} is the dominant one, and that
this contribution is identical (in its pattern of late time decay, and also,
to a certain order of accuracy, in its amplitude) to that
obtained in the shell model.

\subsection* {Contribution from region C} \label{subsecVI1}
In terms of the new integration variables $r_{*}'\equiv (v'-u')/2$
and $t'\equiv (v'+u')/2$, and using Eqs.\ (\ref{eq87}) and (\ref{eq100})
(with $r_{*}\rightarrow \infty$),
the contribution to $\Psi_{1}^{\infty}$ from sources in region {\bf C}
takes the form
\begin{eqnarray} \label{eq104}
\Psi_{1{\rm C}}^{\infty}&=&
-2r_{0}^{-l}\!\sum_{i,j=1}^{l+1} \alpha_{j}\gamma_{li}
\int_{-u/2}^{r_{0}}\!\!\!\!dr_{*}'\int_{-r_{*}'}^{u+r_{*}'}\!\!\!\!\!dt'
E_{i}(u-t'+r_{*}')                                         \nonumber\\
& \times & V(r_{*}')E_{j}(t'+r_{*}'-2r_{0}).
\end{eqnarray}
(recall that in
region $C$, where $r_{*}'<r_{0}$, we have $\delta V=V$ by definition.)

Since $V(r_{*}')\propto \exp[(r_{*}'/2M)]$ for $r_{*}'\rightarrow -\infty$,
one finds that the integrand in the last expression dies off exponentially
in retarded time $u$ anywhere inside the domain of integration.
It is easy to verify that the
integral itself would be exponentially small for large $u/M$.
For example, for any fixed retarded time $u\gg M$ there exist positive
constants $c_{1}$, $c_{2}$, $c_{3}$, $c_{4}$ and $\kappa$, such that
the following upper bound is applicable to the above integral
(in absolute value):
\begin{eqnarray} \label{eq105}
|\Psi_{1{\rm C}}^{\infty}|&\leq&
\sum_{i,j=1}^{l+1} |\alpha_{j}\gamma_{li}| \left\{
c_{1}\int_{-u/2}^{-u/4}dr_{*}' \int_{-r_{*}'}^{u+r_{*}'}\!\!\!dt'\,
V(r_{*}')\right.                                               \nonumber\\
&+&c_{2}\int_{-u/4}^{r_{0}}dr_{*}'
\int_{-r_{*}'}^{\frac{u+r_{*}'}{2}}\!\!\!dt'\,|E_{i}(u-t'+r_{*}')|
                                                               \nonumber\\
&+&\left. c_{3}\int_{-u/4}^{r_{0}}dr_{*}'
       \int_{\frac{u+r_{*}'}{2}}^{u+r_{*}'}\!\!\!dt'\,
       |E_{j}(t'+r_{*}'-2r_{0})|\right\}                        \nonumber\\
&\leq& c_{4}\exp[-\kappa (u/r_{0})].
\end{eqnarray}

We conclude that internal sources of $\Psi_{0}$ (namely sources at
$r_{*}<r_{0}$) give at most an exponentially decaying contribution to the
late time radiation at null infinity.

\subsection*{Contribution from region B} \label{subsecVI2}
By Eqs.\ (\ref{eq79b}) and (\ref{eq94})
(with $r_{*}\rightarrow \infty$), the contribution
from region $B$ to $\Psi_{1}$ at null infinity reads
\begin{eqnarray} \label{eq106}
\Psi_{1{\rm B}}^{\infty}&=&
\sum_{i=1}^{l+1} \sum_{n,j=0}^{l}\!\!\bar{\beta}_{nji}\!\!
\int_{0}^{u+2r_{0}}\!\!\!\!\!\!\!\!\!dv' E_{i}(u-v'+2r_{0})   \nonumber\\
&\times&\int_{u_{0}}^{v'-2r_{0}}\!\!\!\!\!\!\!\!\!du'
\frac{(r_{*}'-r_{0})^{l-j}r_{0}^{j}}{(r_{*}')^{2l-n}}
\delta V(r_{*}') g_{0}^{(n)}(u'),
\end{eqnarray}
in which $\bar{\beta}_{nji}$ are certain constant coefficients,
and the function $g_{0}(u')$ stands for the expressions derived in the
previous chapter for $g_{0}^{I}(u')$ and $g_{0}^{II}(u')$
[Eqs.\ (\ref{eq80a}) and (\ref{eq84}), respectively].
If we now integrate this expression by parts with respect to $v'$, we find
that {\em to the leading order in $M/u$ and in $r_{0}/u$},
\begin{eqnarray} \label{eq107}
\Psi_{1{\rm B}}^{\infty}&=& \sum_{n=0}^{l} \sum_{j=0}^{l}
\tilde{\beta}_{nj}r_{0}
\int_{u_{0}}^{u}du'\frac{(u-u')^{l-j}r_{0}^{j}}{(u-u'+2r_{0})^{2l-n}}
                                                          \nonumber\\
&\times& \delta V(u-u'+2r_{0}) \, g_{0}^{(n)}(u'),
\end{eqnarray}
with $\tilde{\beta}_{nj}$ being some other coefficients.
Now, integrate in parts each of the terms $n$ successive times with respect
to $u'$. The resulting surface terms would all be negligible at large
$u/r_{0}$, since $g_{0}(u')$ dies off exponentially at large retarded time
$u$ [see Eq.\ (\ref{eq84})]. In addition, these surface terms are strictly
compact from below.
We are left with
\begin{eqnarray} \label{eq108}
\Psi_{1{\rm B}}^{\infty}&=& \sum_{n=0}^{l} \sum_{j=0}^{l}(-1)^{n}
\tilde{\beta}_{nj}r_{0}\int_{u_{0}}^{u}du'g_{0}(u') \nonumber\\
&\times& \frac{d^{n}}{du'^{n}}\left[
\frac{(u-u')^{l-j}r_{0}^{j}}{(u-u'+2r_{0})^{2l-n}}\delta V(u-u'+2r_{*})
\right],
\end{eqnarray}
to the leading order in $M/u$ and in $r_{0}/u$.

To continue, we shall have to write $\delta V$ in terms
of the null coordinates. This cannot be done explicitly, since the
function $r(r_{*})$ is implicit. Rather, we shall use the large $r$
expansion
\begin{eqnarray} \label{eq41}
\delta V(r_{*}\geq r_{0})&=&M r_{*}^{-3}[a+b\ln (r_{*}/2M)]
\nonumber\\
&+& \O\left(M^{2}r_{*}^{-4}[\ln (r_{*}/2M)]^{2}\right),
\end{eqnarray}
where $a$ and $b$ are constant coefficients, depending only on $l$ and $M$.
In paper I we argued that (in the framework of the shell model) it
is merely the asymptotic form of the background potential which
affects $\Psi_{1}^{\infty}$ at $u\gg M$. This has been also tested
numerically (see figure 11 in paper I).
We now proceed by assuming that the same is true in the
complete SBH model as well.

With $\delta V$ taken to the leading order in $M/r_{*}$, Eq.\
(\ref{eq108}) takes the form
\begin{eqnarray} \label{eq108a}
\Psi_{1{\rm B}}^{\infty}&=& Mr_{0}\sum_{j=0}^{l} \sum_{m=0}^{l-j}
\int_{u_{0}}^{u}du'g_{0}(u')
\frac{(u-u')^{l-j-m}r_{0}^{j}}{(u-u'+2r_{0})^{2l-m+3}} \nonumber\\
&\times&\left[a_{jm}+b_{jm}ln\left(\frac{u-u'+2r_{0}}{2M}\right)\right],
\end{eqnarray}
to the leading order in $M/u$ and in $r_{0}/u$,
where $a_{jm}$ and $b_{jm}$ are certain constant coefficients.
We observe that, since $g_{0}(u')\sim \exp(-u'/r_{0})$ at large $u'$,
the upper part of the integration over $u'$ (say,
between $u'=\sqrt{uM}$ and $u'=u$) gives a contribution which dies
off exponentially at large $u$. In our approximation we can thus
concentrate on the contribution coming from early retarded times
(say, $u_{0}\leq u'\leq \sqrt{uM}$). At any large enough retarded time
$u$ there exist positive constants $c_{5}$ and $c_{6}$ such that
this early contribution (in absolute value) is bounded from above by
\begin{eqnarray} \label{eq108b}
|\Psi_{1{\rm B}}^{\infty}|&\leq& c_{5}Mr_{0}\sum_{j=0}^{l} \sum_{m=0}^{l-j}
\frac{[(uM)^{1/2}-u_{0}](u-u_{0})^{l-j-m}r_{0}^{j}}
{[u-(uM)^{1/2}+2r_{0}]^{2l-m+3}}                     \nonumber\\
&&\times\left[|a_{jm}|+|b_{jm}|\ln\left(\frac{u-u_{0}+2r_{0}}{2M}\right)
\right]                                                     \nonumber\\
&\leq & c_{6}r_{0}Mu^{-(l+2.5)}\ln (u/M).
\end{eqnarray}
In what follows it will become apparent that this contribution to the
late time radiation at null infinity dies off more rapidly than the
radiation due to scattering in the ``main'' region of sources (region
{\bf A}), which will be shown to be characterized by a $u^{-l-2}$ decay
tail. Therefore, the contribution from region {\bf B} is negligible at
$u\gg M$.

\subsection*{Contribution from region A (the ``main'' region)}
\label{secVI3}
The remaining contribution to calculate is that coming from region {\bf A},
reading
\begin{equation} \label{eq110}
\Psi_{1{\rm A}}^{\infty}(u)=
-\!\int_{u_{0}}^{u}\!\!\!\!du'\!\!\int_{u+2r_{0}}^{\infty}
\!\!\!\!\!\!\!\!\!dv'\,G^{\infty}(u;u',v') \delta V (u',v') \Psi_{0}(u',v').
\end{equation}
To evaluate this expression we first write $\Psi_{0}(u',v')$ in terms of
a function $g_{0}(u')$ [as in Eqs.\ (\ref{eq79}a),(\ref{eq79b})],
then integrate by parts each of the
resulting terms in the r.~h.~s $n$ successive times with respect to $u'$.
Neglecting surface terms, which are all exponentially small at late
time since $g_{0}\sim \exp[-u/r_{0}]$ at large $u$, we obtain
\begin{eqnarray} \label{eq111}
\Psi_{1{\rm A}}^{\infty}(u)&=&-\!\sum_{n=0}^{l} \!(-1)^{n}A_{n}
\int_{u_{0}}^{u}\!\!du'\int_{u+2r_{0}}^{\infty}\!\!\!\!\!\!\!dv'\,g_{0}(u')
                                                         \nonumber\\
&&\times\frac{\partial^{n}}{\partial u'^{n}}\!
\left[\frac{G^{\infty}(u;u',v')\delta V
(u',v')}{(v'-u')^{2l-n}}\right].
\end{eqnarray}

With the explicit form of the Green's function [Eqs.\ (\ref{eq87b})
and (\ref{eq33})], and with $\delta V$ taken
to the leading order in $M/r_{*}$, the last equation takes the form
\begin{eqnarray} \label{eq112}
\Psi_{1{\rm A}}^{\infty}(u)&=& M\sum_{k=0}^{l}\sum_{j=0}^{l-k}
\!\int_{u_{0}}^{u}\!\!\!\!du'\!\!\int_{u+2r_{0}}^{\infty}
\!\!\!\!\!\!\!\!\!\!dv'\,\frac{(u-u')^{l-k-j}(v'-u)^{k}}{(v'-u')^{2l-j+3}}
                                                              \nonumber\\
&&\times \left[\bar{a}_{kj}+\bar{b}_{kj}\ln \left(\frac{v'-u'}{2M}\right)
\right] g_{0}(u').
\end{eqnarray}
where $\bar{a}_{kj}$ and $\bar{b}_{kj}$ are certain
constant coefficients that depend on $l$ and $M$, but {\em not} on $r_{0}$.
Integrating over $v'$, the r.~h.~s of the last equation becomes
\begin{eqnarray} \label{eq113}
M\sum_{k=0}^{l}\sum_{j=0}^{l-k}\sum_{m=0}^{k}
\int_{u_{0}}^{u}\!\!\!\!du'
\frac{(u-u')^{l-k-j}(r_{0})^{k-m}}{(u-u'+2r_{0})^{2l-j-m+2}}
                                                      \times\nonumber\\
\left[\tilde{a}_{kjm}+
\tilde{b}_{kjm}\ln \left(\frac{u-u'+2r_{0}}{2M}\right)\right] g_{0}(u').
\end{eqnarray}
with $\tilde{a}_{kjm}$ and $\tilde{b}_{kjm}$ being yet another constant
coefficients, independent of $r_{0}$.

Now, since $g_{0}(u')$ falls off exponentially at large $u'$, we may
cut off the integration at, say, $u'=(Mu)^{1/2}$ without affecting the
integral to the leading order in $M/u$. Doing so, we observe that the
leading order contribution to this integral comes only from terms
corresponding to $m=k$.
(Note the way the dependence in the parameter $r_{0}$ cancels in
the leading order).

Defining
\begin{equation} \label{eq114}
\int_{u_{0}}^{\sqrt{uM}}g_{0I}(u)du' \simeq
\int_{-\infty}^{\infty} g_{0I}(u)du' \equiv I_{0},
\end{equation}
(were the first equality holds to the leading order in $M/u$, as $g_{0}$
is compact from below and dies off exponentially at large $u$)
we find that to the leading order in $M/u$, in $r_{0}/u$ and in
$u_{0}/u$,
\begin{equation} \label{eq114a}
\Psi_{1}^{\infty}=MI_{0}u^{-l-2}[k_{1}+k_{2}\ln u/M].
\end{equation}
$k_{l}$ and $k_{2}$ are constant coefficients that do {\em not} depend on
$r_{0}$.
The only remaining reference to the value of $r_{0}$ lies within the
integral $I_{0}$.

Since the values of the coefficients $k_{l}$ and $k_{2}$ are independent
of $r_{0}$, then in order to obtain these values one may use
Eq.\ (\ref{eq110}) with whatever value of this parameter (requiring only
that $r_{0}\ll u$). Now, if we take $r_{0}=0$, then
Eq.\ (\ref{eq110}) becomes completely analogous to the
expression for $\Psi_{1}^{\infty}$ {\em in the shell model}
(Eq.\ (37) in paper I).
Comparing these two expressions, we find that the
Green's function $G$ and the potential $\delta V$ appearing in
both integrands are exactly the same.
The two expressions differ only in the form of the
function $\Psi_{0}$, which on both cases is expressed in a similar way
[as in Eq.\ (\ref{eq79a})] in terms of two different functions $g_{0}(u)$.
However, the explicit form of the function $g_{0}(u)$ (as related to the
initial data) has no effect whatsoever on the value of the coefficients
$k_{l}$ and $k_{2}$
and so these coefficients must be the same as in the shell model.
Therefore, comparing Eq.\ (\ref{eq114a}) with Eq.\ (52) of paper I, we
learn that $k_{2}=0$, and that,
to the leading order in $M/u$, in $u_{0}/u$, and in $r_{0}/u$ we
have
\begin{equation} \label{eq52}
\Psi_{1}^{\infty}(u\gg M)=2(-1)^{l+1}(l+1)!MI_{0}\;u^{-l-2}.
\end{equation}
[for the compact initial data setup].

\begin{figure}[htb]
\input{epsf}
\centerline{\epsfysize 6cm  \epsfbox{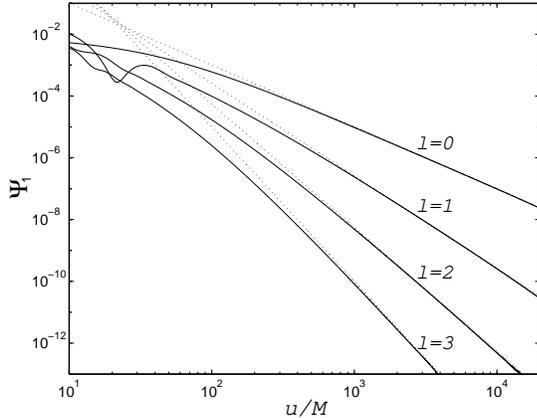}}
\caption{\protect\footnotesize
        Late time tails of $\Psi_{1}$ at null infinity.
        Presented on a log-log scale are numerical results
        obtained for $\Psi_{1}$ in the complete SBH model at
        $v=10^{5}M$ (approximating null-infinity), for the $l=0,1,2,3$
        modes. (For $l=0$ we used the definition of $\Psi_{1}$
        given at the end of section VI).
        Compact initial data for the numerical propagation
        has been specified between $u=-40M$ and $u=-50M$, and the
        parameter $r_{0}$ has been set to $3M$.
        Also shown, for reference, are dotted lines proportional to
        $u^{-l-2}$. The results demonstrate the $u^{-l-2}$ late time
        decay rate predicted by the analytic calculation.}
\label{fig9}
\end{figure}

We conclude that the wave $\Psi_{1}^{\infty}$ has the same
late time behavior in the complete SBH model as it has in the shell model,
namely it dies off as $u^{-l-2}$
provided that the initial pulse is compact.
Numerical calculation of $\Psi_{1}$ in the complete SBH model
agree with this result, as demonstrated in figure \ref{fig9}.

We may similarly obtain the
$u^{-l-1}$ decay characterizing the static initial setup,
by comparing Eq.\ (\ref{eq114a}) with Eq.\ (54) of paper I.
If a static scalar field is present outside the central object up
to some moment before the event horizon forms (no static solution
exist which is well behaved both at the event horizon and at
infinity), we shall have, to the leading order in $M/u$, in $u_{0}/u$,
and in $r_{0}/u$,
\begin{equation} \label{eq52a}
\Psi_{1}^{\infty}(u\gg M)=(-2)^{l+1}\frac{(l!)^{2}}{(2l)!}M\mu\,
u^{-l-1},
\end{equation}
where $\mu$ represents the amplitude of the initial static
field.

It is also instructive to compare the amplitudes of the wave
$\Psi_{1}^{\infty}$ at some fixed $u\gg M$ value on both models
(the shell and the complete SBH), given the same initial
data. As implied by the above discussion, the relative amplitude
is simply given by the ratio
of the integrals $I_{0}\equiv \int_{u_{0}}^{\infty}g_{0}(u')du'$
associated with both models.
Relying on the explicit expressions derived for the functions $g_{0}$
in this paper and in paper I, it can be easily shown that
the two integrals $I_{0}$, associated with the two models, differ
merely by an amount of order $\sim r_{0}/u_{0}$.
Thus, concentrating on the case $|u_{0}|\gg 2M\sim r_{0}$ (in the
context of which our analysis proves effective---see the discussion
in the following section), one observes no difference between the late
time behavior of the wave $\Psi_{1}$ at null infinity on both models.
This result is accurate to the
leading order in $M/u$, in $u_{0}/u$ and in $r_{0}/u_{0}$.
In particular, we may conclude that at late time the wave
$\Psi_{1}^{\infty}$ has no reference (in our approximation) to the
value of neither the radius of the shell R (in
the shell model) nor the parameter $r_{0}$ (used in the complete
SBH model).
This result is consistent with the assumption that details of
spacetime structure at small $r$ values are not manifested in the form of
the late time radiation.

\subsubsection*{The monopole case ($l=0$)}

It was pointed out while introducing the iterative expansion, that
a scheme based on that expansion fails to handle correctly the
case $l=0$. In the monopole case it is straightforward to find
that a Green's function defined as in sec.\ \ref{secV} is {\em
constant} (a unity) throughout the whole range of evolution. It
then follows that the wave $\Psi_{1}$ is constant at late time,
resulting in the divergence of higher terms of the expansion
($\Psi_{2},\Psi_{3},\ldots$). Looking for the cause of this
failure, we notice that in the cases $l>0$ it is the centrifugal
potential barrier that ``cuts off'' the Green's function and
confines it (for late retarded time evaluation points, $u\gg M$)
mainly to late retarded times ($u<u'<v$). In the shell model (see
paper I) it was the presence of the center of symmetry which
effectively acted as a potential barrier for the Green's function
even in the monopole case, where no centrifugal potential exists.
This is why the iterative expansion applied in the framework of
the shell model proved to be equally effective for all modes of
the radiation.

To analyze the case $l=0$ in the complete SBH model, one is thus
led to try a different iterative expansion, defined
such that the Green's function is subject to an appropriate
potential barrier, as for the modes with $l>0$.
One technically simple possibility it to take
\begin{equation} \label{eq53}
V^{l=0}_{0}\equiv M^{-1}\delta(r_{*}).
\end{equation}
We then define the iterative expansion as in Eqs.\ (\ref{eq77a}),
(\ref{eq78}),(\ref{eq78a}), with the `new' potential
$V_{0}^{l=0}$.
With this definition, we find that $\Psi_{0}^{l=0}=\Gamma(u)$, and
that the Green's function (at an evaluation point $u,v$ with
$(v-u)/2=r_{*} >0$) is given by
\begin{equation} \label{eq54}
\left\{
       \begin{array}{l}
       G(u'\leq u,u\leq v'\leq v)=1                \\
       G(u'\leq v' \leq u)=\exp[(v'-u)/M]          \\
       G(v'\leq u'\leq u)=\exp[(u'-u)/M]
       \end{array}
\right. .
\end{equation}
(The three regions indicated in this equation are those labeled 'A','B',
and 'C', respectively, in Fig.\ (\ref{fig7}), when setting $r_{0}=0$).
A simple calculation [based on Eq.\ (\ref{eq102})] then shows that
at null infinity $\Psi_{1}^{l=0}$ is given by Eqs.\ (\ref{eq52}) and
(\ref{eq52a}) (with $l=0$). These equations are therefore valid for all
modes $l$.

\section{Higher terms of the iterative expansion} \label{secVII}
The Green's function technique applied thus far provides a formal way to
calculate each of the terms $\Psi_{N}$ one at a time, in an inductive manner.
We shall not, however, calculate further terms of the expansion, but
rather we will refer to the results of the analysis in paper I.
Regarding the shell model, strong indications were given that
\begin{itemize}
\item The dominant contribution to $\Psi_{N}^{\infty}(u\gg M$), for all
$N\geq 1$, is only due to sources of $\Psi_{N-1}$ at large
distances.
\item All terms $\Psi_{N}$ of the iterative expansion (excluding
$\Psi_{0}$) seem to share the same late time pattern of decay at null
infinity, namely a $u^{-l-2}$ inverse power-law tail (for compact initial
pulse), or a $u^{-l-1}$ tail (for static initial field).
\item If the initial pulse is confined to large distance, $|u_{0}|\gg M$,
then the iterative sum converges at null infinity rather efficiently to the
`complete' scalar wave $\Psi$ at late retarded time $u$.
\item Moreover, in this case ($|u_{0}|\gg M$), the
scalar wave $\Psi$ is well approximated by merely $\Psi_{1}$
(with corrections smaller by order $M/u_{0}$).
\end{itemize}

Now, it is reasonable to assume that all four of the above
results are also valid in the complete SBH model.
For, in both models (the SBH and the shell models)
spacetime structure at large distance is the same,
and, as argued in paper I,
it is the large distance region whose structure is relevant
in determining the late time form of the waves at null-infinity.
(This conclusion has been demonstrated in the framework of the shell
model by explicit analytic calculation of $\Psi_{1}$
and $\Psi_{2}$ at null-infinity. Physical explanation was given
in the concluding section of paper I.
For the complete SBH model, a demonstration is provided by the explicit
analytic calculation of $\Psi_{1}^{\infty}$ in the preceding section).
Actually, we need only to assume that the first of the four results
indicated above holds in the complete SBH model. Then, the same
reasoning used in section VII of paper I in deriving Eq.\ (72),
leads us immediately to realize that $\Psi_{N}\sim u^{-l-2}$ in the
SBH model as well. Also, a completely analogous analysis to that
applied in section VIII of paper I shows that the third and forth
of the above results hold in the complete SBH model as well.

Numerical analysis of the complete SBH model firmly supports the
above arguments, showing that all four
results are indeed valid in the complete SBH model.
In what follows we present some examples of these numerical experiments.

\begin{figure}[htb]
\input{epsf}
\centerline{\epsfysize 6cm \epsfbox{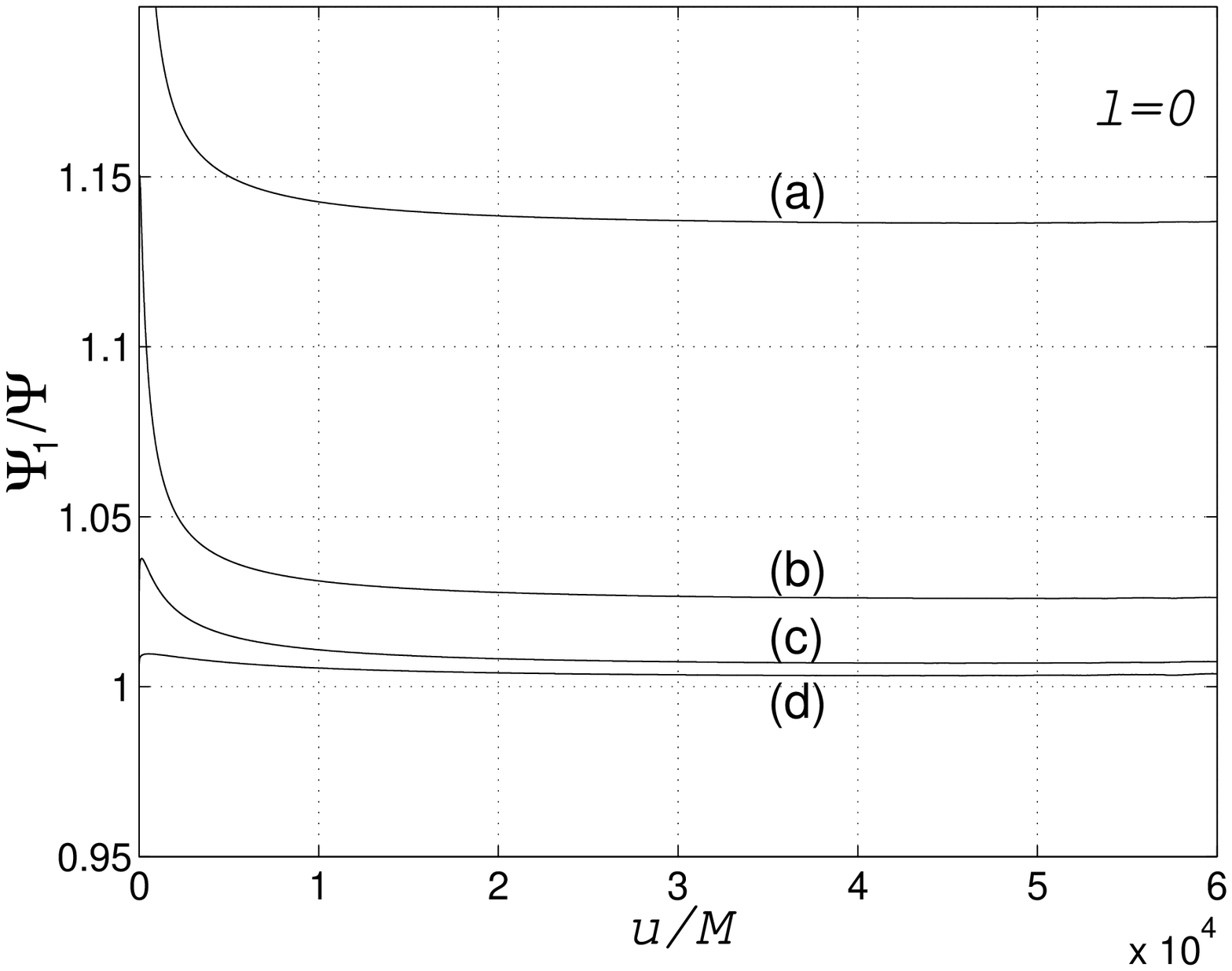}}
\centerline{\epsfysize 6cm \epsfbox{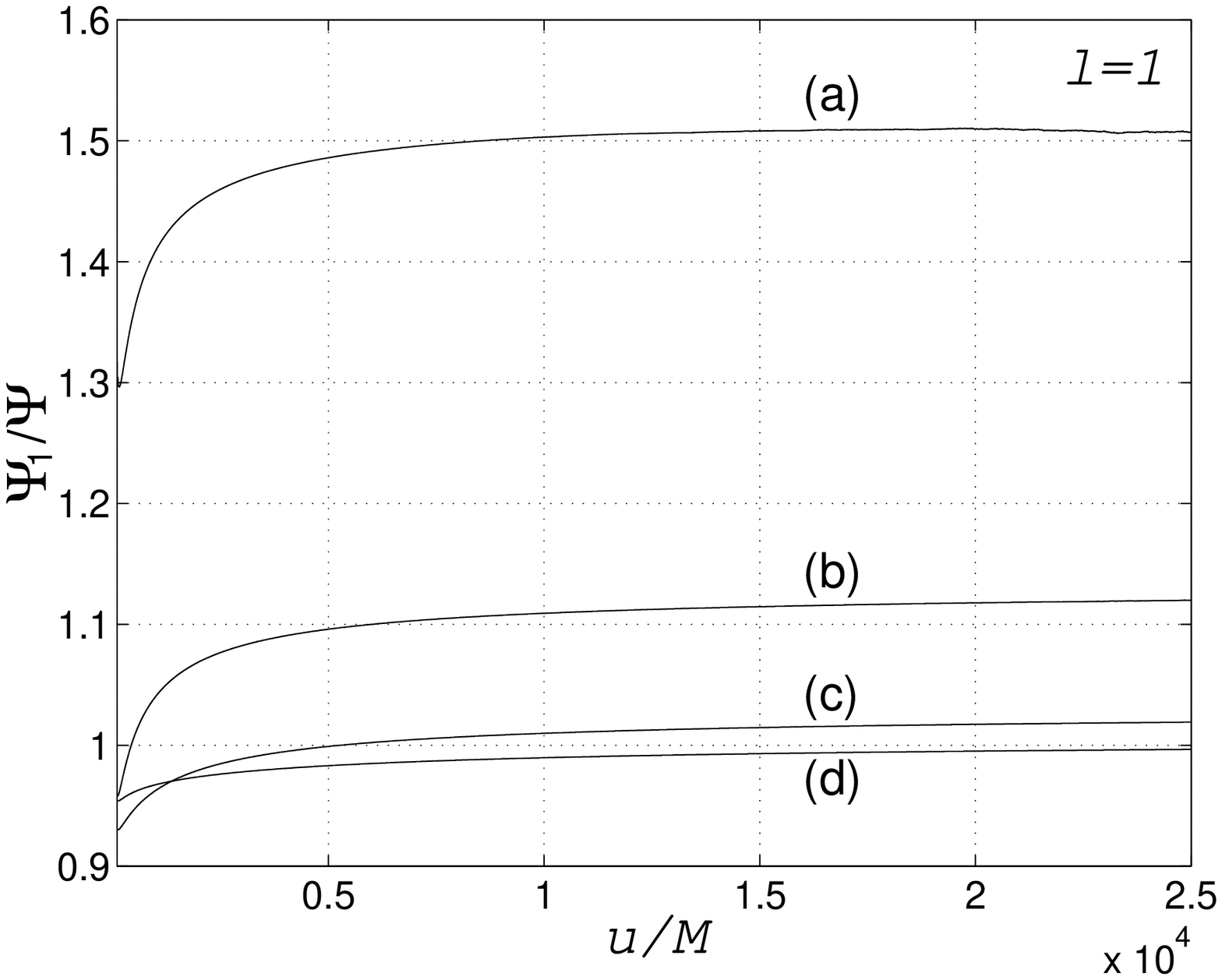}}
\caption{\protect\footnotesize
        $\Psi_{1}$ approximates the ``complete'' wave $\Psi$ at
        null infinity at late time, provided that the parameter $|u_{0}|/M$
        is chosen large. This is demonstrated numerically for the
        monopole ($l=0$) and the dipole ($l=1$) radiation by comparing the
        ratios $\Psi_{1}/\Psi$ at null infinity (approximated by $v=10^{5}M$)
        for the various values (a) $u_{0}=-40M$, (b) $u_{0}=-200M$, (c)
        $u_{0}=-1000M$, and (d) $u_{0}=-5000M$.
        The parameter $r_{0}$ is set to $3M$.}
\label{fig10}
\end{figure}

Figure \ref{fig10} presents the ratio $\Psi_{1}^{\infty}/\Psi^{\infty}$,
calculated numerically for the monopole and the dipole modes, for various
values of the parameter $u_{0}$.
(The ``complete'' wave $\Psi$ has been obtained by a
direct numerical solution of Eq.\ (\ref{eq4})).
The results demonstrate that (like in the shell model) as
$|u_{0}|/M$ is set larger, $\Psi_{1}$ becomes a better approximation to the
``complete'' wave $\Psi$ at null infinity at late time.

\begin{figure}[htb]
\input{epsf}
\centerline{\epsfysize 6cm  \epsfbox{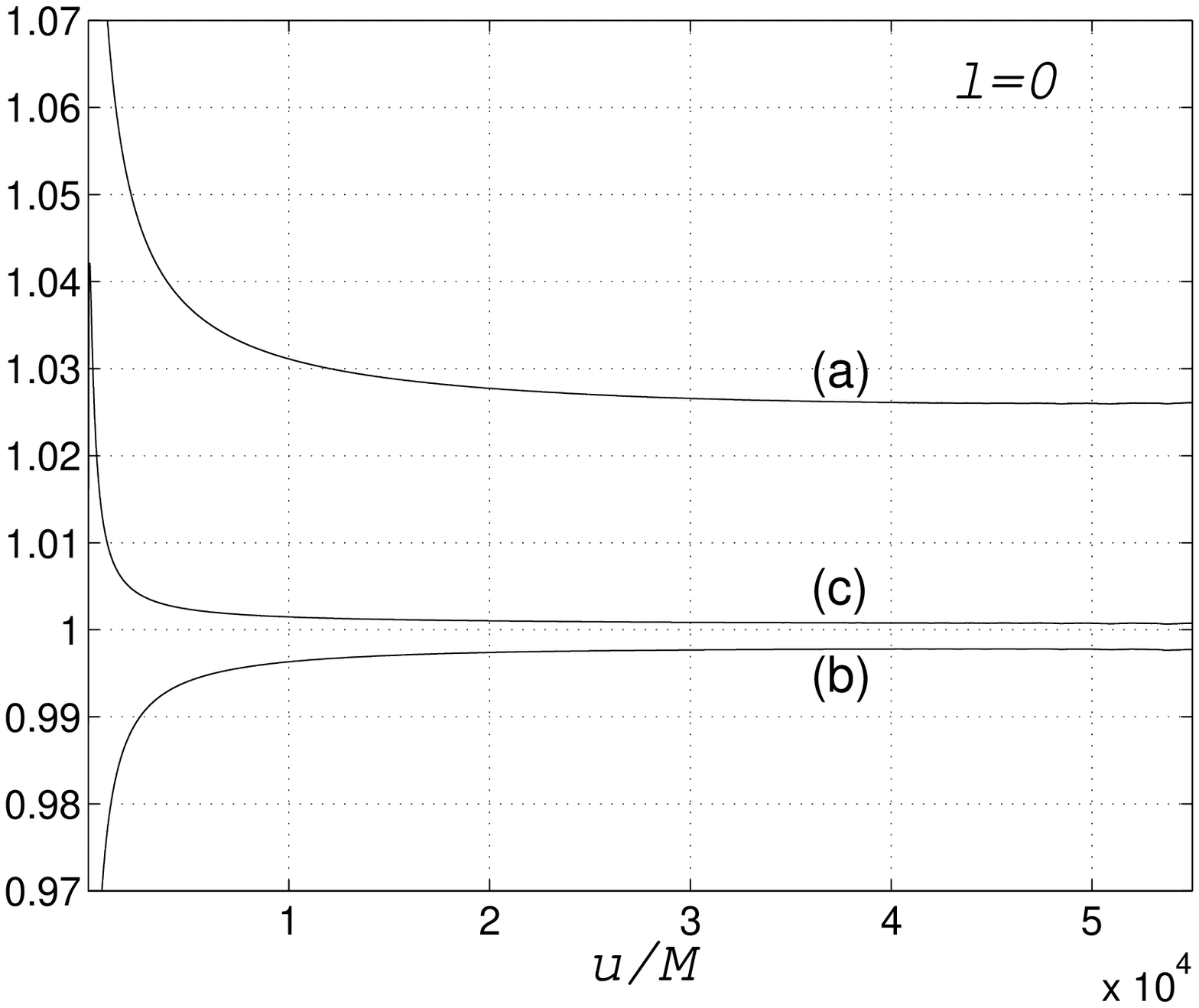}}
\centerline{\epsfysize 6cm  \epsfbox{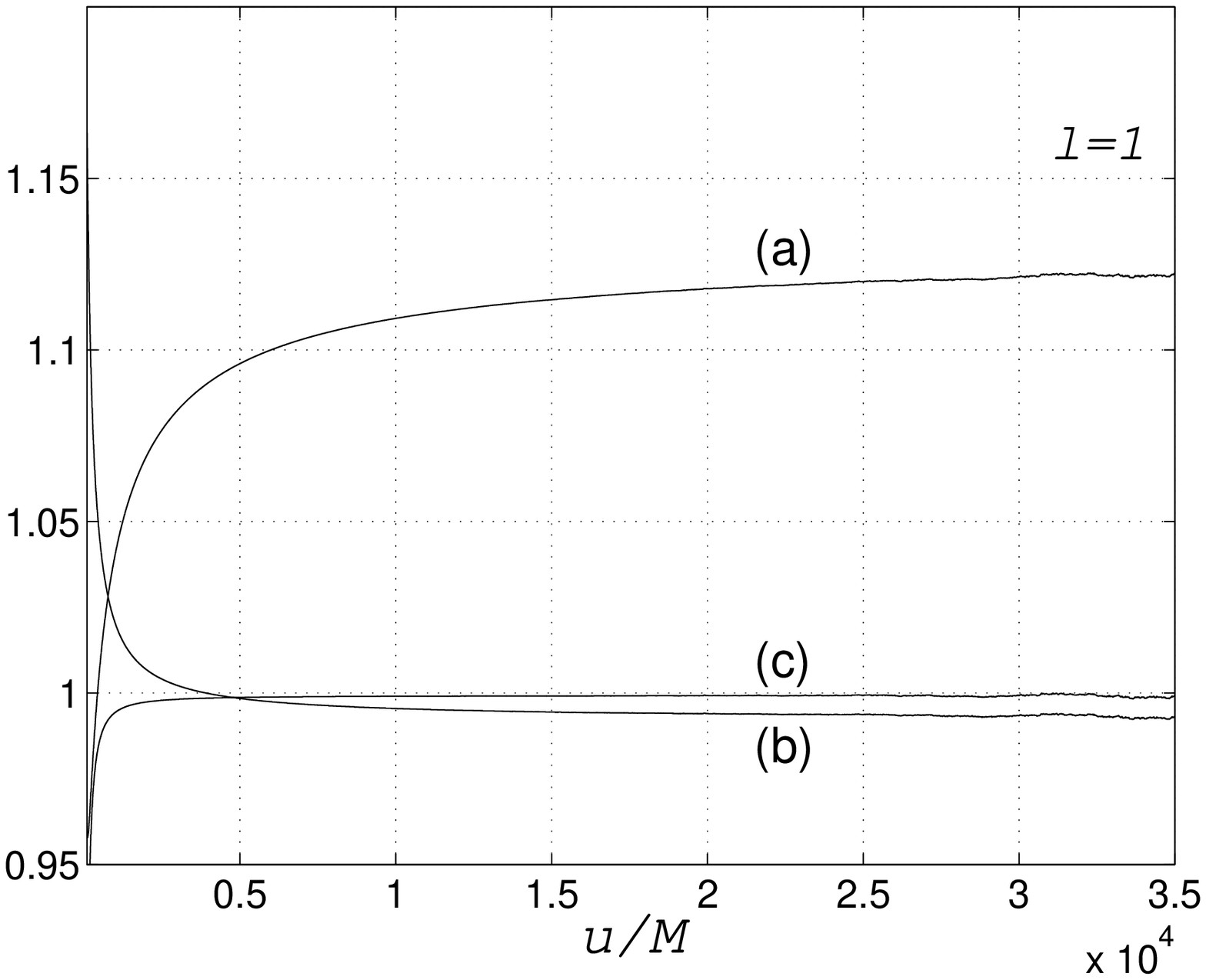}}
\centerline{\epsfysize 6cm  \epsfbox{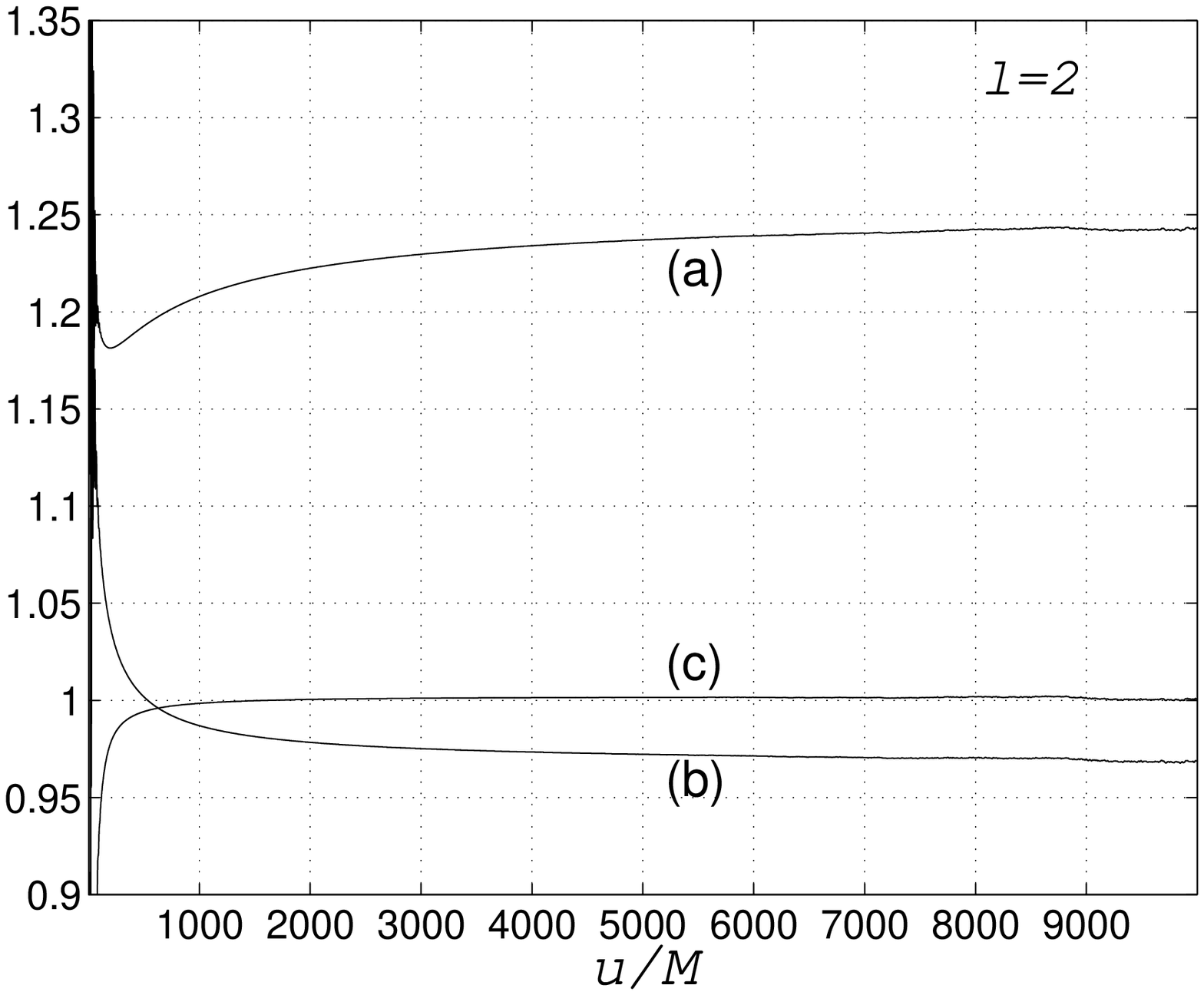}}
\caption{\protect\footnotesize
        Numerical indications for convergence of the iterative scheme
        at null infinity.
        Presented (on a linear scale) are the ratios (a)
        $\Psi_{1}/\Psi$, (b) $\frac{\Psi_{1}+\Psi_{2}}{\Psi}$, and (c)
        $\frac{\Psi_{1}+\Psi_{2}+\Psi_{3}}{\Psi}$ for the sample values
        $l=0,1,2$.
        The other parameters are set to $r_{0}=3$, $u_{0}=-200M$ and
        $v=10^{5}M$ (approximating null infinity).
        The results suggest a rather efficient convergence of the iterative
        expansion for large $|u_{0}|/M$ values at large retarded time $u$.}
\label{fig11}
\end{figure}

In figure \ref{fig11} it is demonstrated numerically (for
$l=0,1,2$) that the iterative series applied in this paper seems to
converge rather efficiently for a large $|u_{0}|/M$ value.

\section{Tails at constant radius: the late time expansion}
\label{secVIII}
So far we were discussing an analytic technique enabling the
calculation of the late time behavior of the scalar field at null
infinity.
In this section we apply a different, local, analysis to study the
late time behavior of the wave along any $r=const$ world-line outside
the black hole, and along the event horizon. Using this method we
will be able to derive a simple analytic expression for the
field, consistent with an inverse power-law decay, and accurate to the
leading order in $M/t$ (or in $M/v$ along the
horizon). However, this expression will inhere two undetermined parameters
(one for the power-law index and the other for the amplitude).
We shall deduce these parameters by matching our late time solution
at null infinity to the form derived in the previous sections using the
iterative scheme.
In that respect, the iterative scheme shall
prove to be an essential key for the construction of a complete
late time description of the wave behavior anywhere outside the black hole.
The purpose of this section is to introduce and apply the local method.

We make the assumption that at late time, the Klein--Gordon
scalar wave [priorly separated in terms of the spherical harmonics as in
Eq.\ (\ref{eq2})] admits the expansion
\begin{equation} \label{eqR1}
\phi^{l}(r,v)=\sum_{k=0}^{\infty}F^{l}_{k}(r)v^{-k_{0}-k},
\end{equation}
in which the number $k_{0}$ and the set of functions $F_{k}(r)$ are yet
to be determined.
Substituting this expression in the Klein--Gordon equation (\ref{eq3}) and
collecting terms of common $v$-power, the partial equation is thereby
converted to an infinite set of {\it ordinary} coupled equations for the
unknown functions $F_{k}(r)$,
\begin{eqnarray} \label{eqR2}
r^2\left(1-\frac{2M}{r}\right)F_{k}'' +2(r-M)F_{k}'-l(l+1)F_{k}=\nonumber\\
2(k_{0}+k-1)(r^2F_{k-1}'+rF_{k-1})
\end{eqnarray}
(for $k\geq 0$), where a prime denotes differentiation with respect to $r$,
and where we have set $F_{-1}\equiv 0$.
This set of equations exhibit only a `weak' coupling, in the sense that
each of the functions $F_{k}$ depends only on its preceding function
$F_{k-1}$, with $F_{0}$ obeying a closed homogeneous equation. This
hierarchy allows one to treat each of these equations one by one, in an
iterative way. In this procedure, each of the functions $F_{k}$
(with $k\geq 1$) satisfies a closed second-order inhomogeneous equation.

We proceed as follows: First, we show that there exists a solution
$\phi^{l}$ of the form (\ref{eqR1}), which is regular anywhere outside the
black hole, in particular at the event horizon and at null
infinity.
Then, with the aid of our previous results at null infinity,
we deduce the late time behavior of the scalar waves as detected
by a static observer at any constant radius. In particular, we obtain
the late time form of the waves along the event horizon.

We define a new dimensionless radial coordinate
\begin{equation} \label{eqR3}
\rho \equiv \frac{r-M}{M},
\end{equation}
which varies monotonically from the event horizon ($\rho=1$) up to
space-like infinity ($\rho=\infty$).
In terms of the new variable, Eq.\ (\ref{eqR2}) takes the form
\begin{equation} \label{eqR4}
(\rho^{2}-1)F_{k}''+2\rho F_{k}'-l(l+1)F_{k}=D_{k}(\rho)[F_{k-1}],
\end{equation}
in which a prime now denotes differentiation with respect to $\rho$, and
where $D(\rho)$ is the differential operator
\begin{equation} \label{eqR5}
D_{k}(\rho)=2M(k_{0}+k-1)(\rho+1)[(\rho+1)\partial_{\rho}+1].
\end{equation}
(Note that $D_{k}(\rho)$ depends on $k_{0}$, but is independent of $l$).
For $k=0$, the r.~h.~s of Eqs.\ (\ref{eqR4}) vanishes.

We would like to construct solutions $F_{k}$ to Eqs.\ (\ref{eqR4}) such
that $\phi$ would be regular both at the event horizon and at infinity.
To allow $\phi$ to be regular at the horizon
(where the $v$ coordinate takes finite values),
we require all functions $F_{k}(\rho)$ to be regular at
$\rho=1$\footnote{
Had we apply a $1/t$ expansion of $\phi$ instead of the $1/v$ expansion
(\ref{eqR1}), all functions $F_{k}$ had to diverge at the horizon
(where $t=\infty$) to assure regularity of the scalar wave there.
For that reason, the $1/v$ expansion seems more plausible from the
technical point of view.}.
We will now show that such regular solutions $F_{k}(\rho)$ do
exist\footnote{In Mathematical terminology, Eq.\ (\ref{eqR4}) possesses a
``regular singular point'' at $\rho =1$ (the event horizon)---see, for
example, \cite{Wayland57}. For this case, standard mathematical
theory tells that series solution for $F_{0}$ and for $F_{1}$ can
always be constructed near $\rho =1$. However, for $k\geq 2$ the
source term in Eq.\ (\ref{eqR4}) appears to involve logarithmic functions,
for which case standard theory gives no clear rules.}.

We shall construct the functions $F_{k}$ in an iterative way, starting
with $F_{0}$. For $k=0$, Eq.\ (\ref{eqR4}) is homogeneous, and its general
solution is given by $F_{0}^{l}=a_{0}P_{l}(\rho)+b_{0}Q_{l}(\rho)$,
where $a_{0}$ and $b_{0}$ are arbitrary parameters, $P_{l}(\rho)$ is the
Legendre polynomial of order $l$, and $Q_{l}(\rho)$ is the Legendre function
of the second kind, of order $l$.
The polynomials $P_{l}(\rho)$ are, of course, finite at the
event horizon ($\rho=1$) and divergent (as $\rho^{l}$) at
$\rho\rightarrow \infty$.
Conversely, The functions $Q_{l}(\rho)$ diverge at the event horizon and
vanish
(as $\rho^{-l-1}$) at $\rho\rightarrow \infty$. Regularity of $F_{0}$
at the event horizon therefore requires that $b=0$, hence we obtain
\begin{equation} \label{eqR6}
F_{0}^{l}(\rho)=a_{0}P_{l}(\rho).
\end{equation}

Now consider Eq.\ (\ref{eqR4}) for a general function $F_{k}$
(with $k\geq 1$).
The general solutions to the inhomogeneous equations read
\begin{eqnarray} \label{eqR7}
F_{k}^{l}(\rho)&=&a_{k}P_{l}(\rho)+b_{k}Q_{l}(\rho)           \nonumber\\
 & &\mbox{} + P_{l}(\rho)\int_{1}^{\rho}\frac{Q_{l}(\rho')D_{k}(\rho')
     [F_{k-1}(\rho')]}{(\rho'^{2}-1)W(\rho')}d\rho'           \nonumber\\
 & &\mbox{} - Q_{l}(\rho)\int_{1}^{\rho}\frac{P_{l}(\rho')D_{k}(\rho')
     [F_{k-1}(\rho')]}{(\rho'^{2}-1)W(\rho')}d\rho',
\end{eqnarray}
where $a_{k}$ and $b_{k}$ are arbitrary parameters, and
\begin{equation} \label{eqR8}
W\equiv P_{l}'Q_{l}-P_{l}Q_{l}'=(\rho^{2}-1)^{-1}
\end{equation}
is the Wronskian.
Using the relation
\begin{equation} \label{eqR9}
Q_{l}(\rho)=P_{l}(\rho)\int_{\rho}^{\infty}\frac{W(\rho')}{P_{l}^{2}(\rho')}
d\rho'
\end{equation}
and integrating Eq.\ (\ref{eqR7}) by parts, we can then obtain
\begin{eqnarray} \label{eqR10}
\lefteqn{F_{k}^{l}(\rho)  =  a_{k}P_{l}(\rho)+b_{k}Q_{l}(\rho)+ }\nonumber\\
& & P_{l}(\rho)\int_{1}^{\rho}\!\!\!d\rho'\frac{W(\rho')}
   {P_{l}^{2}(\rho')}
   \int_{1}^{\rho'}\!\!\!d\rho''P_{l}(\rho'')D_{k}(\rho'')[F_{k-1}(\rho'')].
\end{eqnarray}

We now show by mathematical induction that with $b_{k}=0$ (for all $k$),
the functions $F_{k}(\rho)$ are all {\em analytic} at the event horizon.
The first function, $F_{0}$, is
analytic at $\rho=1$ by Eq.\ (\ref{eqR6}). Now, following the inductive
procedure, assume that $F_{k-1}$ is
analytic at $\rho=1$ for some $k\geq 1$. Then,
$D_{k}(\rho'')[F_{k-1}(\rho'')]$ is analytic at $\rho''=1$,
hence the integrand of the $\rho''$ integration in Eq.\ (\ref{eqR10}) is
analytic at that point. We thus find that the integral over $\rho''$
can be written in the form $(\rho'-1)\bar{f}(\rho')$, where
$\bar{f}(\rho)$ is some function which is
analytic at the horizon (this can be shown by expanding the integrand in
a Taylor series near $\rho''=1$, where it is analytic).
Since the polynomials $P_{l}$ have no real zeros is the range
$\rho\geq 1$ \cite{GradRyz} , and $W$ diverges as $(\rho'-1)^{-1}$ at
the horizon, we conclude that the whole integrand of the $\rho'$
integration is analytic at $\rho'=1$, and therefore that the integral
over $\rho'$ must be analytic as well.
Hence the solutions $F_{k}$, defined in an inductive way by
Eq.\ (\ref{eqR10}), with $b_{k}=0$ for all $k\geq 1$, are all analytic
at the horizon.

By this we have shown that the wave equation admits solutions
$\phi$ of the form (\ref{eqR1}), which are analytic at the event horizon.
The most general of these solutions contains an infinite number of
free parameters, one for each power of $1/v$

We do not know yet the value of the power index $k_{0}$, appearing in the
expansion (\ref{eqR1}). To obtain this value we shall now evaluate
$\phi$ at null infinity.
By this mean we will be able to (i) show that the form of $\phi$ at
null infinity is consistent with the results of our iterative analysis
(in particular, that $\phi$ is regular there); and (ii) deduce the value
of $k_{0}$ by comparing the results arising from the two
independent schemes.

We start by showing, using mathematical induction, that the functions
$F_{k}$ all have the asymptotic form
\begin{equation} \label{eqR11}
F_{k}(\rho\rightarrow \infty)\sim a_{0}c_{k}\rho^{l+k},
\end{equation}
in which $c_{k}$ are certain constant coefficients, other then zero and yet
to be determined.
Here and henceforth, the form $f(x)\sim cx^{n}$ (where $c$ is some constant)
means that $\lim_{x\rightarrow \infty}[f(x)/x^{n}]= c$.
It appears most convenient to prove Eq.\ (\ref{eqR11})
by first showing that\footnote
{This is not valid when $l=0$ and $k=0$, for which case $dF_{k}/d\rho=0$.
However, the rest of the analysis does not change.}
\begin{equation} \label{eqR12}
\frac{dF_{k}}{d\rho}(\rho\rightarrow \infty)\sim a_{0}(l+k)c_{k}\rho^{l+k-1}.
\end{equation}
Then, Eq.\ (\ref{eqR11}) is implied.

The form (\ref{eqR12}) obviously applies for $F_{0}$ [given in
Eq.\ (\ref{eqR6})], with
\begin{equation} \label{eqR13}
c_{0}= \frac{(2l+1)!!}{l!(2l+1)},
\end{equation}
which is just the coefficient of $\rho^{l}$ in the polynomial $P_{l}(\rho)$.

Following the inductive procedure, we now assume that
Eq.\ (\ref{eqR12}) applies for some $k\geq 0$,
and show that this leads to
$dF_{k+1}/d\rho\sim (l+k+1)a_{0}c_{k+1}\rho^{l+k}$.
Our assumption necessarily implies that $F_{k}\sim a_{0}c_{k}\rho^{l+k}$.
Hence, by Eq.\ (\ref{eqR5}) we have
\begin{equation} \label{eqR14}
D_{k+1}F_{k}\sim 2Ma_{0}c_{k}(k_{0}+k)(l+k+1)\rho^{l+k+1}.
\end{equation}
Consequently, for the integration over $\rho''$ in Eq.\ (\ref{eqR10})
we obtain the asymptotic form
\begin{equation} \label{eqR15}
\sim 2Ma_{0}c_{k}c_{0}\frac{(k_{0}+k)(l+k+1)}{(2l+k+2)}(\rho')^{2l+k+2}.
\end{equation}
It then follows from Eq.\ (\ref{eqR10}) that
\begin{equation} \label{eqR16}
F_{k+1}\sim a_{0}c_{k+1}\rho^{l+k+1},
\end{equation}
with
\begin{equation} \label{eqR17}
c_{k+1}=2Mc_{k}\frac{(k_{0}+k)(l+k+1)}{(2l+k+2)(k+1)}.
\end{equation}
Finally, differentiating Eq.\ (\ref{eqR16}), we get
$dF_{k+1}/d\rho\sim (l+k+1)a_{0}c_{k+1}\rho^{l+k}$,
which establishes the inductive proof of Eq.\ (\ref{eqR12}).

We have thereby shown that the functions $F_{k}$ all admit the
asymptotic form (\ref{eqR11}), with the coefficients $c_{k}$ given by the
recursive formula (\ref{eqR17}), supplemented by Eq.\ (\ref{eqR13}).
We obtain, in conclusion,
\begin{equation} \label{eqR18}
F_{k}\sim a_{0}\alpha_{l}(2)^{k}C_{k}M^{-l}r_{*}^{l+k}
\end{equation}
for $r_{*}\rightarrow \infty$, where we have explicitly used the fact that
$\rho\sim r/M\sim r_{*}/M$, and where
\begin{equation} \label{eqR19}
C_{k}=\frac{(k_{0}+k-1)!(l+k)!}{(2l+k+1)!(k)!},
\end{equation}
and
\begin{equation} \label{eqR20}
\alpha_{l}=\frac{(2l)!(2l+1)!!}{(l!)^{2}(k_{0}-1)!}.
\end{equation}
(The coefficients $C_{k}$ and $\alpha_{l}$ are not to be confused
with the coefficients appearing in sec.\ (\ref{secIV})).

Eq.\ (\ref{eqR18}) describes the form of the functions $F_{k}$
to the leading order in $r/M$, which is sufficient for our purpose:
matching $\phi$ at null infinity.
We comment, however, that a full series expression for the
functions $F_{k}$ at large $r$ can be obtained as well. It has the form
\begin{equation} \label{eqR21}
F_{k}(r)=r^{k}\sum_{j=0}^{k}(r/M)^{l-j}H_{kj}(M/r)[\ln (r/M)]^{j},
\end{equation}
where $H_{kj}(M/r)$ are Taylor series.
This form can be verified by substituting it into Eq.\
({\ref{eqR2}), then constructing explicit recursive formulae for the
coefficients of each of the various series $H_{kj}$.
Note that to the leading order in $r/M$ no logarithmic terms are involved,
and the form (\ref{eqR18}) is recovered.

To obtain $\phi$ at null infinity, we insert Eq.\ (\ref{eqR18}) into the
expansion ({\ref{eqR1}). We get (to the leading order in $r/M$),
\begin{equation} \label{eqR22}
\phi=a_{0}\alpha_{l}(2M)^{-l}v^{l-k_{0}}\sum_{k=0}^{\infty}C_{k}
\left(1-\frac{u}{v}\right)^{l+k}.
\end{equation}

To evaluate the power series, we write it in terms of a generating
function,
\begin{equation} \label{eqR23}
\sum_{k=0}^{\infty}C_{k}q^{k}=
q^{-2l-1}\frac{d^{k_{0}-2l-2}}{dq^{k_{0}-2l-2}}\left[q^{k_{0}-1}
\frac{d^{l}}{dq^{l}}\left(\frac{q^{l}}{1-q}\right)\right],
\end{equation}
which is valid for $|q|< 1$. In this expression the derivatives might be
of negative orders, in which case integrations are implied.
If we now make the substitution $q=\left(1-\frac{u}{v}\right)$, keeping
just the leading order in $u/v$, we find that
\begin{eqnarray} \label{eqR24}
\sum_{k=0}^{\infty}C_{k}\left(1-\frac{u}{v}\right)^{k}  =
\frac{d^{k_{0}-l-2}}{dq^{k_{0}-l-2}}\left(\frac{1}{1-q}\right)= \nonumber\\
(k_{0}-l-2)!\left(\frac{v}{u}\right)^{k_{0}-l-1}.
\end{eqnarray}
Since $\Psi\equiv r\phi =v\phi$ to the leading order in $u/v$, we finally
obtain
\begin{equation} \label{eqR25}
\Psi^{\infty}(u)=a_{0}\alpha_{l}(k_{0}-l-2)!(2M)^{-l}
u^{-(k_{0}-l-1)},
\end{equation}
where $\Psi^{\infty}(u)$ stands for the wave $\Psi$ evaluated at null
infinity.

The value of both the power index $k_{0}$ and the yet-free parameter
$a_{0}$ can now be specified by comparing the last result to the results
arising from our iterative scheme, Eqs.\ (\ref{eq52}) and (\ref{eq52a}).
This comparison yields (assuming that $|u_{0}|\gg M$),
\begin{equation} \label{eqR26}
k_{0}=\left\{ \begin{array}{ll}
                2l+3  & \mbox{\rm no initial static field}\\
                2l+2  & \mbox{\rm initial static field}
                \end{array}
       \right.,
\end{equation}
and
\begin{equation} \label{eqR27}
a_{0}=\left\{ \begin{array}{ll}
                \frac{(l!)^{2}(2l+2)!}{(2l)!(2l+1)!!}(-2M)^{l+1}I_{0}
                 & \ \mbox{\rm no initial static field.}\\
                \frac{(l!)^{3}(2l+1)}{2(2l)!(2l+1)!!}(-4M)^{l+1}\mu
                 & \ \mbox{\rm initial static field}
                \end{array}
       \right.,
\end{equation}
where the integral $I_{0}$ [defined in Eq.\ (\ref{eq114})] is
directly related to the initial data via Eq.\ (\ref{eq80a}),
and $\mu$ is the amplitude
of the initial static field (when present).

Provided with an exact expression for $F_{0}(r)$ and with the value of
$k_{0}$, we are now in a position to write the form of the scalar field at
any finite value of $r$, at very late time. By Eq.\ (\ref{eqR1}) we have,
{\em to the leading order in $M/t$ and in $M/u_{0}$},
\begin{equation} \label{eqR28}
\Psi=\frac{a_{0}}{t^{2l+2,3}}\,r\,P_{l}\left(\frac{r-M}{M}\right)
, \ \ \ (t\gg|r_{*}|) ;
\end{equation}
and {\em to the leading order in $M/v$ and in $M/u_{0}$},
\begin{equation} \label{eqR29}
\Psi=\frac{2Ma_{0}}{v^{2l+2,3}} ,
\mbox{\ \ \ \ at the event horizon},
\end{equation}
where the two values of powers correspond to the cases where an initial
static field is or is not present, respectively.

We should emphasize here that the above results [Eqs. (\ref{eqR28}) and
(\ref{eqR29})] apply towards time-like infinity at
{\em any} value of $r$, establishing [together with the results at
null infinity, Eqs.\ (\ref{eq52}), (\ref{eq52a})] a {\em complete}
picture of the late time behavior outside the black hole.
To the best of our knowledge, such a result has never been obtained
previously.
(For example, in \cite{Price72}, \cite{Gundlach94I}, and \cite{Leaver86},
analytic expressions were derived only for the asymptotic
domains $r_{*}\gg M$ and $r_{*}\ll M$.)

\section{concluding remarks} \label{secIX}

In this paper, and in the paper preceding it, we have tested the
applicability of a new analytic scheme for the calculation of the late time
behavior of fields outside black holes.
It was demonstrated, considering the simple model of scalar waves
outside a SBH, that a simple expansion of the field near time-like
infinity can be used in order to construct a late time solution consistent
with a power-law decay anywhere outside the black hole.
However, the actual index of the power-law, as well as the
amplitude coefficient of the wave (as related to the initial data)
could not be determined merely by this local analysis.
This information could be obtained only by a full integration of the
2-dimensional initial value problem for the wave evolution, technically
enabled by the introduction and application of the iterative
procedure.

Thus, by applying both the iterative scheme and the late time
expansion, we were able to obtain an analytic expression for the
scalar field in Schwarzschild, accurate to the leading order
in $M/t$ (or in $M/u$ at null infinity, or in $M/v$ at the
event horizon) and holding anywhere outside the black hole.
The expression calculated is explicitly  related
[via Eqs.\ (\ref{eq80a}), (\ref{eq114}), and (\ref{eqR27})]
to the form of arbitrary initial data specified at large distance
(our approximate solution has corrections of order $M/u_{0}$).

Of course, the main justification for the introduction of the new
approach should rely on its applicability to more realistic models
of wave evolution, for which no other analytic approaches have been
proposed. In what follows we mention some possible applications
of our calculation scheme, which include the
analysis of scalar fields outside Kerr black holes, the
analysis of gravitational perturbations in Kerr, and the extension
of our analysis to the interior of black holes.

The most interesting application of the new scheme concerns
rotating black holes.
As already mentioned in the introduction, this generalization
is the prime motivation behind the presentation of our approach, since
realistic stellar objects (and black holes) generically posses angular
momentum. The generalization of our analysis to the
case of scalar waves propagating in the exterior of a Kerr black hole
shall be presented in a forthcoming paper.
In brief, the basic idea behind this generalization is to express
the lack of spherical symmetry in Kerr spacetime in
terms of {\em interactions} between the various modes of spherical
harmonics. The resulting interaction terms coupling the field
equations for the various modes (these terms are expected to be
small, in a sense, at late time)
are then to be treated using our iterative technique.
Applying an iterative decomposition basically similar to that used
in the spherically-symmetric models, those interaction terms
become source terms in the resulting hierarchy of wave equations.
The mathematical treatment of these equations is then similar,
in principle, to that applied in the spherically-symmetric cases.
This provides the late time form of each of the modes at null
infinity. Then, a generalization of the late time
expansion method (based on the same interaction-between-modes
approach) provides the late time behavior of the field
anywhere outside the Kerr black hole, in particular along its
event horizon.
The details of both parts of the analysis in Kerr shall be given
in \cite{Ours} (see also \cite{Barack97}).

Obviously, the scalar model discussed so far is just a simplified
analogue to the physical problem concerning the dynamics of gravitational
perturbations. The plausibility of the scalar model stems from the
remarkable resemblance of the underlying mathematical formulation,
between this model and realistic models of gravitational waves (as was
already realized in \cite{Price72}, for example, for the SBH case).
Equations governing metric perturbations of the SBH were derived by
Regge and Wheeler \cite{Reg57} (for axial
perturbations) and by Zerilli \cite{Zer70} (for polar perturbations).
Both equations can be put in the same form as the scalar field
equation (\ref{eq4}), where this time the wave function represents certain
linear combinations of entities characterizing the metric perturbation. In
both Regge-Wheeler's and Zerilli's equations one also finds that the
effective potential is similar in shape to that of the scalar model
(Eq.(6) and figure 1).
This suggests that the problem of gravitational waves propagation
in the SBH geometry may be treated using the same iterative
scheme applied for the scalar model.

We do not have similar separable equations for metric
perturbations of Kerr black holes. Rather, a second approach, based on the
Newman--Penrose tetrad formalism, was used by Teukolsky \cite{Teukolsky72}
to derive separable wave equations governing perturbations of the Weyl
scalars.
[An exhaustive discussion of field equations for gravitational
perturbations is given in \cite{Chandra83}, pp.\ 174-182 (for Schwarzschild)
and pp.\ 430-443 (for Kerr)].
Separation of Teukolsky's equation is only possible in the
frequency domain (namely by first separating the wave to its
Fourier modes).
To apply our iterative approach, however, the time dependence should rather
be kept in the master Teukolsky equation.
Instead, the master perturbation equation is to be treated in the way
outlined above, namely, by considering interactions between the various
modes of spherical harmonics.
By now we have first indications that this approach is indeed applicable
to gravitational perturbations (in the tetrad formalism).
We intend to study this subject more deeply in the near future.

Finally, we mention the possibility of extending our analysis to internal
perturbations of black holes. Recently, Ori used the technique of
{\em late time expansion} (basically similar to the method presented in
sec.\ \ref{secVIII} of the present paper), to explore the late time
behavior of scalar fields {\em inside} charged \cite{Ori97} and rotating
\cite{Ori98} black holes.
In this analysis, boundary conditions for the wave evolution were assumed
on the event horizon (in the form of an inverse power-law in $v$), and the
asymptotic late time ($t\gg M$) behavior of the wave was deduced inside
the black hole, up to the inner horizon. This provided a tool for exploring
the nature of the inner horizon singularity.
With the results of the external analysis (generalized to charged
and rotating black holes), a connection may be established between
the form of the wave at the inner horizon to its form at null
infinity, which, in turn, using the iterative scheme, can be derived as
explicitly related to the form of arbitrary initial data outside
the black hole. That would allow one, given initial data outside the
black hole (at large distance), to deduce the late time form of
the wave at the inner horizon (including its accurate amplitude
coefficient) without any assertion about the boundary conditions.

One may also think of a more rigorous and coherent scheme, which
includes the simultaneous analysis of both internal and external
perturbations, in the framework of a generalized late time expansion.
This generalization becomes natural when
applying an expansion of the form (\ref{eqR1}), as the coordinates
$v$ and $r$ are both regular through the event horizon. Then a full
treatment of both internal and external evolution is possible by
following basically the same steps as described in sec.\ \ref{secVIII},
this time allowing the $r$ coordinate to take its full range of values.

\section*{acknowledgments}
The author wishes to express his indebtedness to Professor A.\ Ori
for his guidance throughout the execution of this research and for
countless helpful discussions.




\begin{references}

\bibitem[*]{Email}{E.\ mail: leor@techunix.technion.ac.il}

\bibitem{Price72} R. H. Price, Phys.\ Rev.\ D {\bf 5}, 2419 (1972);
{\bf 5}, 2439 (1972).

\bibitem{Gundlach94I} C. Gundlach, R. H. Price, and J. Pullin, Phys.\ Rev.\
D {\bf 49}, 883 (1994).

\bibitem{Winicour94} R. G\'{o}mez, J. Winicour, and B. G. Schmidt,
Phys.\ Rev.\ D {\bf 49}, 2828 (1994).

\bibitem{Leaver86} E. Leaver, J. Math.\ Phys.\ (N.Y.) {\bf 27}, 1238 (1986);
Phys.\ Rev.\ D {\bf 34}, 384 (1986).

\bibitem{Andersson97} N. Andersson, Phys.\ Rev.\ D {\bf 55}, 468 (1997).

\bibitem{Ching95} E. S. C. Ching, P. T. Leung, W. M. Suen, and K. Young,
Phys.\ Rev.\ Lett.\ {\bf 74}, 2414 (1995).

\bibitem{Burko97} L. M. Burko, A. Ori, Phys.\ Rev.\ D {\bf 56}, 7820 (1997).

\bibitem{Gundlach94II} C. Gundlach, R. H. Price, and J. Pullin, Phys.\ Rev.\
D {\bf 49}, 890 (1994).

\bibitem{Brady97} P. R. Brady, C. M. Chambers, W. Krivan, and P. Laguna,
Phys.\ Rev.\ D {\bf 55}, 7538 (1997).

\bibitem{Krivan96} W. Krivan, P. Laguna, and P. Papadopoulos,
Phys.\ Rev.\ D {\bf 54}, 4728 (1996).

\bibitem{Krivan97} W. Krivan, P. Laguna, P. Papadopoulos, and N. Andersson,
Phys.\ Rev.\ D {\bf 56}, 3395 (1997).

\bibitem{Ours} L. Barack, A. Ori, {\em Late time tails in Kerr geometry},
in preparation.

\bibitem{Barack97} L. Barack in {\em Internal structure of black holes and
spacetime singularities}, Volume XIII of the Israel Physical Society,
Edited by L. M. Burko and A. Ori, (Institute of Physics, Bristol, 1997).

\bibitem{Friedlander75} F.\ G.\ Friedlander, {\em The wave equation on a
curved spacetime} (Cambridge University Press, Cambridge, 1975),
section 5.4 (see in particular theorems 5.4.1 and 5.4.2).

\bibitem{Wayland57} H. Wayland, {\em Differential equations
applied in science and engineering} (D. Van Nostrand Company,
princeton, New Jersey, 1957), pp.\ 131---146.

\bibitem{GradRyz} I.\ S.\ Gradshteyn, I.\ M.\ Ryzhik,
{\em Tables of integrals, series and products}, \S8.782.

\bibitem{Reg57} T. Regge and J. A. Wheeler, Phys. Rev. {\bf 108}, 1063
                  (1957).

\bibitem{Zer70} F. J. Zerilli, Phys.\ Rev.\ D{\bf 2}, 2141 (1970).

\bibitem{Teukolsky72} S. A. Teukolsky, Phys.\ Rev.\ Lett.\ {\bf 29}, 1114
(1972)

\bibitem{Chandra83} S. Chandrasekhar, {\em The Mathematical Theory
of Black Holes} (Oxford University Press, New York, 1983).

\bibitem{Ori97} A. Ori, Phys.\ Rev.\ D{\bf 55}, 4860 (1997)

\bibitem{Ori98} A. Ori, Phys.\ Rev.\ D, in press.

\end{references}
\end{document}